\documentclass[fleqn,usenatbib]{mnras}

\usepackage{newtxtext,newtxmath}

\usepackage[T1]{fontenc}

\usepackage{graphicx}	
\usepackage{amsmath}	
\usepackage{gensymb}
\usepackage{threeparttable}

\usepackage{caption}
\usepackage{subcaption}

\graphicspath{{./}{figs/}}

\let\oldtextbf\textbf
\renewcommand{\textbf}[1]{\oldtextbf{\boldmath #1}}

\title[Growing evidence for the neutrino-blazar connection]{Growing evidence for high-energy neutrinos originating in radio blazars}

\author[Plavin et al.]{
A. V. Plavin,$^{1}$\thanks{E-mail: alexander@plav.in}
Y. Y. Kovalev,$^{2,1,3}$
Yu. A. Kovalev$^{1}$,
and S. V. Troitsky$^{4,5}$
\\
$^{1}$Lebedev Physical Institute of the Russian Academy of Sciences, Leninsky prospekt 53, 119991 Moscow, Russia\\
$^{2}$Max-Planck-Institut f\"ur Radioastronomie, Auf dem H\"ugel~69, 53121 Bonn, Germany\\
$^{3}$Moscow Institute of Physics and Technology, Institutsky per.~9, 141700 Dolgoprudny, Russia\\
$^{4}$Institute for Nuclear Research of the Russian Academy of Sciences, 60th October Anniversary prospect~7a, 117312 Moscow, Russia
\\
$^{5}$Lomonosov Moscow State University, 1-2 Leninskie Gory,  Moscow 119991, Russia}

\date{Accepted 2023 May 03. Received 2023 April 12; in original form 2022 November 17}

\pubyear{2023}

\begin{document}
\label{firstpage}
\pagerange{\pageref{firstpage}--\pageref{lastpage}}
\maketitle

\defcitealias{neutradio1}{P20}
\defcitealias{neutradio2}{P21}


\begin{abstract}
Evidence for bright radio blazars being high-energy neutrino sources was found in recent years. However, specifics of how and where these particles get produced still need to be determined. 
In this paper, we add 14 new IceCube events from 2020-2022 to update our analysis of the neutrino-blazars connection.
We test and refine earlier findings by utilising the total of 71 track-like high-energy IceCube events from 2009-2022. We correlate them with the complete sample of 3412 extragalactic radio sources selected by their compact radio emission.
We demonstrate that neutrinos are statistically associated with radio-bright blazars with a post-trial $p$-value of $3\cdot10^{-4}$.
In addition to this statistical study, we confirm previous individual neutrino-blazar associations, find and discuss several new ones. Notably, PKS~1741$-$038 was selected earlier and had a second neutrino detected from its direction in 2022; PKS~0735+168 has experienced a major flare across the whole electromagnetic spectrum coincidently with a neutrino arrival from that direction in 2021.
\end{abstract}

\begin{keywords}
neutrinos --
galaxies: active --
galaxies: jets --
quasars: general --
radio continuum: galaxies
\end{keywords}



\section{Introduction} \label{sec:intro}

IceCube has been detecting astrophysical neutrinos of TeV to PeV energies for more than a decade \citep{IceCubeFirst26}, and other neutrino observatories have recently confirmed these results \citep{ANTARES2019ICRC,Baikal-Neutrino2022a,2022arXiv221109447B}. However, the origins of these particles are still not fully determined: this is challenging because of a strong atmospheric background and the relatively poor angular resolution of current neutrino telescopes. See, e.g. the \cite{2022PhyU...64.1261T} review for a summary of recent findings and challenges.

Despite these difficulties, multiple associations of neutrinos with celestial objects were made. The first significant association of an individual point source was the TXS~0506+056 blazar with a high-energy neutrino detected in 2017 \citep{IceCubeTXSgamma}. Later, tens of bright radio blazars were associated with neutrinos (\citealt{neutradio1}, hereafter in the paper, we refer to it as \citetalias{neutradio1}; \citealt{neutradio2}, hereafter \citetalias{neutradio2}) and more papers have evaluated the neutrino-blazar connection \citep[e.g.][]{Resconi2020,2021A&A...650A..83H,2021PhRvD.103l3018Z,RadioPS,Illuminati2021,Franckowiak2022,2022ApJ...933L..43B}. General constraints on the neutrino source population also remain consistent with blazars generating a sizeable fraction of the total flux \citep[e.g.][]{2022arXiv221004930A}. Currently, the general neutrino-blazar connection can only be demonstrated on a statistical basis, given the background and resolution of neutrino detectors.

In this paper, we provide an up-to-date analysis of the neutrino-radio blazar associations. We test our earlier findings and predictions from \citetalias{neutradio1} on the basis of new IceCube events. We find confirmations for previous associations and suggest new ones. In addition to a statistical analysis, we discuss notable individual neutrino-blazar associations and put them into a wider context of active galaxies with highly relativistic jets pointing at us.

\begin{figure*}
\includegraphics[width=0.95\linewidth]{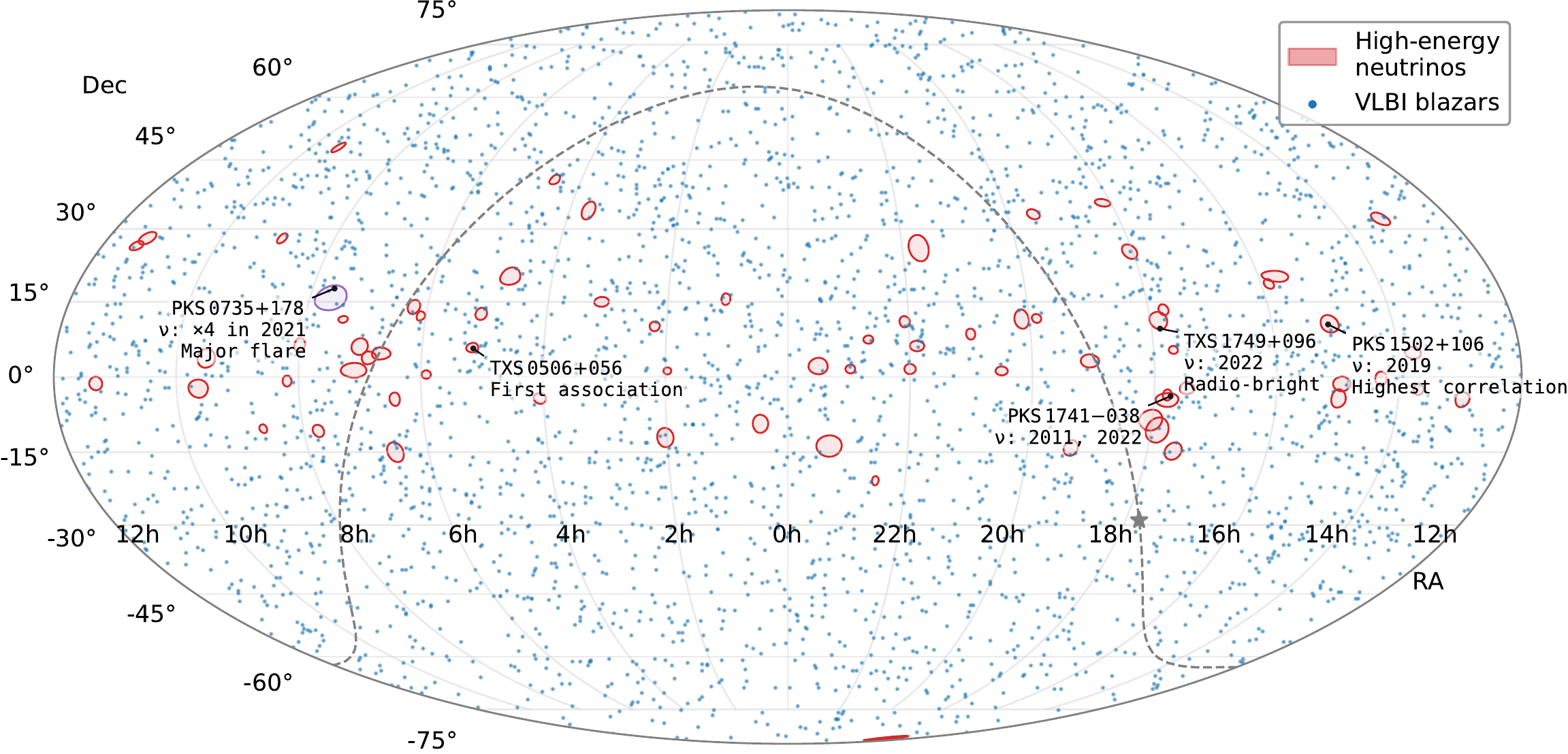}
\caption{
\label{f:skymap}
Directions of the IceCube high-energy events as red ellipses (\autoref{s:data_icecube}) and radio bright blazars as blue points (\autoref{s:data_vlbi}). The ellipses represent the event error regions enlarged by $\Delta=0.45\degree$ to account for systematic errors, see \autoref{s:stat_flux} and \autoref{f:pvals} for details. The magenta ellipse corresponds to the event discussed in \autoref{s:J0738}, which is not within the \autoref{s:data_icecube} sample. The five blazars discussed in \autoref{s:individual} are labelled here.
}
\end{figure*}

\section{Data} \label{s:data}
\subsection{IceCube Events} \label{s:data_icecube}

We follow previous works (\citetalias{neutradio1}; \citealt{2021A&A...650A..83H}) and utilise the same selection criteria for the highest-energy neutrino detections at IceCube: track morphology of the event, neutrino energy $E\geq200$~TeV and the reported 90\% error region area below $10$~sq.~deg.\ on the celestial sphere. Events from 2009 to 2019 were collected in \citetalias{neutradio1,neutradio2}, and we use them without any modifications for the most direct comparison. There are 56 of those from \citetalias{neutradio1} and one more event from 2012 with a reported energy of exactly $E = 200$~TeV, which we missed in \citetalias{neutradio1} but included in \citetalias{neutradio2}, see also \citet{2021A&A...650A..83H}. Events before 2019 include extremely high energy (EHE) detections that pass our selection criteria by construction and other events we collected from publicly available lists, see \citetalias{neutradio1} or references in \autoref{t:events}.
We gathered 14 more recent events from the beginning of 2020 to 1~November~2022, passing the same criteria from IceCube GCN alerts.
Together, this set of 71 tracks includes alerts and alert-like events from 2009 to 2022.
They all are listed in \autoref{t:events} and are visualised in \autoref{f:skymap}. These events allow us to perform a direct test of our early findings presented in \citetalias{neutradio1}, see \autoref{s:stat_flux}.

\begin{table}
\caption{High-energy IceCube neutrino detections used in our analysis.}
\label{t:events}
\centering
\begin{tabular}{ccrcrcc}
\hline\hline
Date & $E$ & \multicolumn{2}{c}{RA} & \multicolumn{2}{c}{DEC} & Reference \\
 & (TeV) & \multicolumn{2}{c}{($\degr$)} & \multicolumn{2}{c}{($\degr$)} & \\
(1) & (2) & (3) & (4) & (5) & (6) & (7) \\
\hline
2020-06-15 & 496 & 142.95 & $_{-1.45}^{+1.18}$ &     3.66 & $_{-1.06}^{+1.19}$ & GCN\,27950\\[0.1cm]
2020-09-26 & 670 &  96.46 & $_{-0.55}^{+0.73}$ &  $-4.33$ & $_{-0.76}^{+0.61}$ & GCN\,28504\\[0.1cm]
2020-10-07 & 683 & 265.17 & $_{-0.52}^{+0.52}$ &     5.34 & $_{-0.23}^{+0.32}$ & GCN\,28575\\[0.1cm]
2020-11-14 & 214 & 105.25 & $_{-1.12}^{+1.28}$ &     6.05 & $_{-0.95}^{+0.95}$ & GCN\,28887\\[0.1cm]
2020-11-30 & 203 &  30.54 & $_{-1.31}^{+1.13}$ & $-12.10$ & $_{-1.13}^{+1.15}$ & GCN\,28969\\
\hline
\end{tabular}
\begin{tablenotes}
\item Notes:
Columns are as follows: (1) is the IceCube event date; (2) is the best estimate of the neutrino energy, when available; (3), (4), (5) and (6) are the event equatorial coordinates with their reported errors; (7) reference to the paper or GCN with details of the detection.
\item
The set of 71 IceCube events selected according to our criteria, see \autoref{s:data_icecube} for details. The detections until 2019 are directly transferred from  \citetalias{neutradio1} (56 events) and \citetalias{neutradio2} (one event) and included here to make reproducing the most up-to-date analysis convenient. Only a portion of this table with five new events is shown here to demonstrate its form and content. A machine-readable version of the full table is available online.
\end{tablenotes}
\end{table}

\subsection{Blazars: VLBI Data} \label{s:data_vlbi}

We base our analysis on the complete sample of blazars observed by VLBI, following \citetalias{neutradio1,neutradio2}. Positions and parsec-scale flux densities of those objects are compiled in the Radio Fundamental Catalogue\footnote{\url{http://astrogeo.org/sol/rfc/rfc_2022b/}} (RFC). The VLBI observations were performed at the 8~GHz band, including geodetic VLBI \citep{2009JGeod..83..859P,2012A&A...544A..34P,2012ApJ...758...84P}, the Very Long Baseline Array (VLBA) calibrator surveys (VCS; \citealt{2002ApJS..141...13B,2003AJ....126.2562F,2005AJ....129.1163P,2006AJ....131.1872P,2007AJ....133.1236K,2008AJ....136..580P,r:wfcs,2016AJ....151..154G}), other 8~GHz global VLBI, VLBA, EVN (the European VLBI Network) and LBA (the Australian Long Baseline Array) observations 
\citep{2011AJ....142...35P,2011AJ....142..105P,2011MNRAS.414.2528P,2012MNRAS.419.1097P,2013AJ....146....5P,2015ApJS..217....4S,2017ApJS..230...13S,2019MNRAS.485...88P}.
The complete flux density limited sample of VLBI-selected blazars consists of 3412 objects with historic average 8-GHz flux density $S^\mathrm{VLBI}_\mathrm{8\,GHz}\geq150$~mJy integrated over their VLBI images. This blazar sample is used in the analysis throughout the paper, the objects are shown as dots in \autoref{f:skymap}.
While the RFC collects VLBI flux density results at multiple radio frequencies, we chose 8~GHz due to superior completeness characteristics of the sample at this band.

\section{Studying samples of blazars and high energy neutrinos} \label{s:sample_analysis}
\subsection{Statistical analysis} \label{s:stat_flux}

We conduct an independent test of the results and predictions given in \citetalias{neutradio1}. Specifically, we perform the same calculations utilising IceCube events detected since the publication of that paper in 2020. There are 14 high-energy events in this time period (\autoref{s:data_icecube}), and we cannot expect to reach a high statistical significance focusing on them alone; for comparison, there were 56 events included in the \citetalias{neutradio1} analysis. The main goal of using these new events is to check earlier results for consistency and reproducibility. Furthermore, we repeat the same statistical calculation using the whole sample of 71 events. This lets us obtain the most up-to-date significance estimates of the blazar~--- high energy neutrino correlation.

Here, we briefly motivate and describe the statistical procedure; for more details on the algorithm, see Appendix~\ref{a:statdetails}. Compact parsec-scale radio emission from blazars is an indicator of energetic relativistic processes happening in the jet and of the Doppler boosting effect \citep{1997ARA&A..35..607Z,2019ARA&A..57..467B,2021ApJ...923...67H}. This emission is measured by VLBI. RFC provides historical averaged flux density $S^\mathrm{VLBI}_\mathrm{8\,GHz}$ integrated over 8~GHz VLBI images, and we use it in the analysis (\autoref{s:data_vlbi}).
We directly follow \citetalias{neutradio1}, and take the geometric average of $S^\mathrm{VLBI}_\mathrm{8\,GHz}$ of all blazars within the event error regions as the test statistic. 
The event error regions are taken as IceCube uncertainties enlarged by a certain value $\Delta$, interpreted as a measure of the systematic error; see its determination and discussion in the next paragraph and in Appendix~\ref{a:statdetails}.
Then, a Monte-Carlo method is employed to test if the average is significantly higher than could arise by chance: the $p$-value is the fraction of random realizations yielding more extreme test statistic values than the real data.

\begin{figure}
\includegraphics[width=1.0\columnwidth]{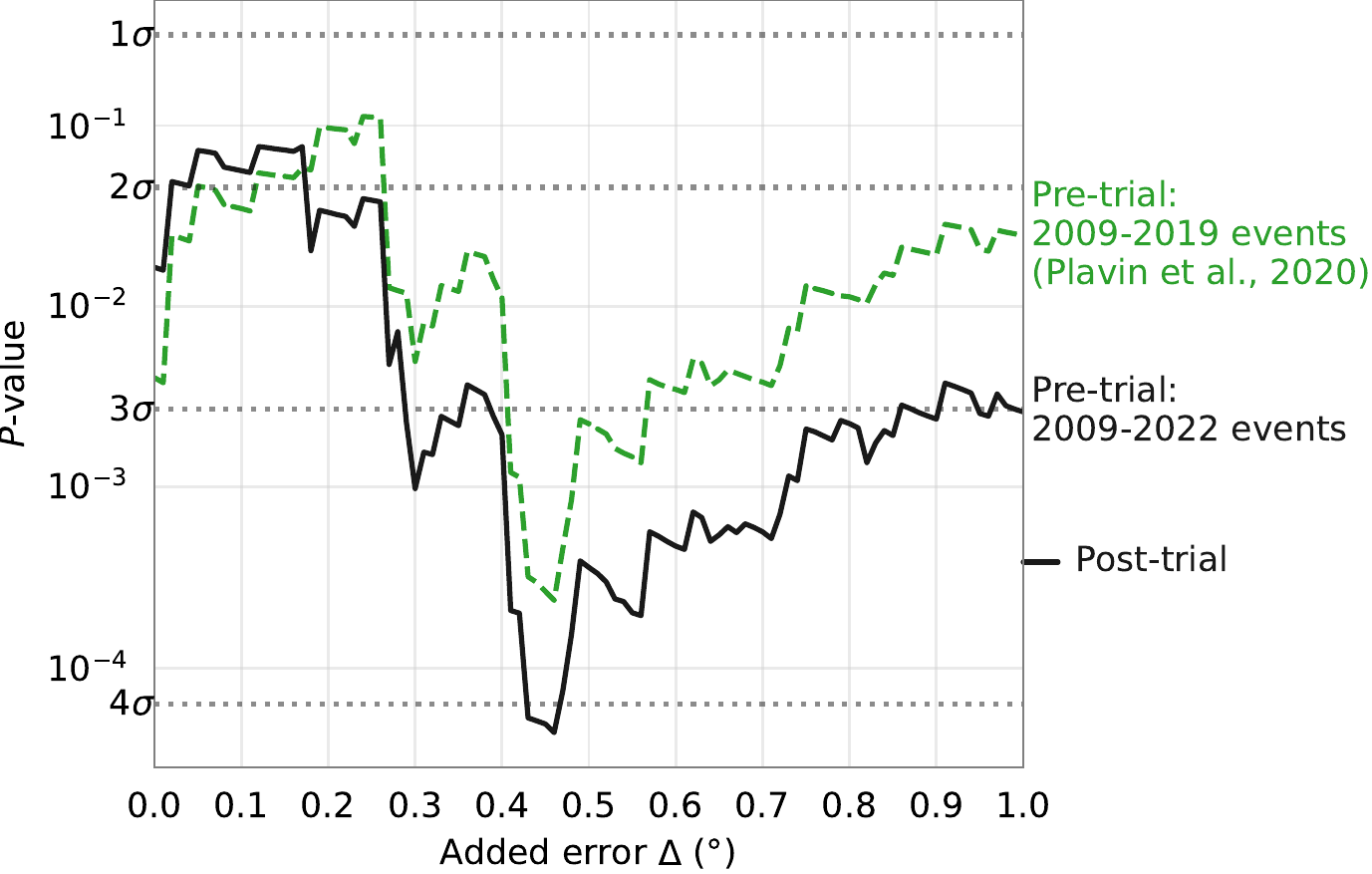}
\caption{$P$-values for bright VLBI blazars (\autoref{s:data_vlbi}) being spatially correlated with IceCube high-energy neutrinos (\autoref{s:data_icecube}). The horizontal axis marks the added error $\Delta$ defined in \autoref{s:stat_flux}. The two curves correspond to the sample of 56 neutrino detections until 2019 \citepalias{neutradio1} and to the most up-to-date sample of 71 events used in this work. Minima of both curves are attained at $\Delta=0.45\degree$, which is the value we use for the subsequent statistical evaluations in this paper. The final post-trial $p$-value for all 2009--2022 events is $p=3\cdot10^{-4}$, marked in this plot; analogously, for 2009--2019 events, the post-trial $p=2\cdot10^{-3}$.}
\label{f:pvals}
\end{figure}

This analysis has a single free parameter $\Delta$: the value added to the original uncertainty regions reported for IceCube events. We, again, interpret $\Delta$ as an indirect measure of the IceCube systematic error; accounting for systematics in more direct ways with simulations remains hard or impossible for now \citep{2021arXiv210708670L}. The optimal magnitude of $\Delta$ was statistically estimated to be $0.5\degree$ in \citetalias{neutradio1}.
Multiple groups have tested the results presented in \citetalias{neutradio1}, directly or indirectly, discussing those tests with us. Following these discussions, we corrected a minor inconsistency in treating $\Delta$ in the original code, and performed the parameter scan with a step of $0.01\degree$ in $\Delta$. As a result, our pre-trial $p$-value estimates accuracy has increased: see the line corresponding to 2009~--~2019 events in \autoref{f:pvals}. The optimal value of $\Delta$ slightly changed and became $0.45\degree$. The post-trial $p$-value based on the events until 2019, originally reported in \citetalias{neutradio1}, does not change and remains 0.2\%.

We repeat the same statistical analysis with the 14 new events since 2020, keeping $\Delta = 0.45\degree$ determined using the earlier 2009~--~2019 events only. This avoids the multiple comparisons issues, yielding a $p$-value of 6\% that does not require any trial corrections. This is a completely independent check of our earlier predictions based on 2009~--~2019 events. This again points towards a spatial correlation of high-energy neutrinos and radio-bright blazars. Taking into account that the new 2020~--~2022 sample of events is thrice smaller compared to \citetalias{neutradio1}, the $p$-value of 6\% is in agreement with the findings reported in that paper. Selected individual associations with recent IceCube events are highlighted and discussed in \autoref{s:individual}.

\begin{table}
\caption{Positional associations between the blazars and high-energy IceCube events.}
\label{t:assocs}
\centering
\begin{tabular}{cccccc}
\hline\hline
Event Date & \multicolumn{2}{c}{Blazar name}  & $z$   & $S^{\text{VLBI}}$ & $d$       \\
         & J2000                    & B1950 &       & (Jy)              & ($\degr$) \\
(1)      & (2)                      & (3)   & (4)   & (5)               & (6)       \\
\hline
2010-10-09 & J2200$+$1030 & 2157$+$102 &      & 0.18 & 1.20\\
2010-10-09 & J2203$+$1007 & 2201$+$098 & 1.00 & 0.21 & 0.99\\
2010-11-13 & J1855$+$0251 & 1853$+$027 &      & 0.16 & 2.07\\
2010-11-13 & J1858$+$0313 & 1855$+$031 &      & 0.67 & 1.44\\
2011-07-14 & J0432$+$4138 & 0429$+$415 & 1.02 & 1.36 & 1.34\\
2011-09-30 & J1743$-$0350 & 1741$-$038 & 1.05 & 4.02 & 0.75\\
2012-05-15 & J1310$+$3220 & 1308$+$326 & 1.00 & 1.90 & 1.02\\
2012-05-15 & J1310$+$3233 & 1308$+$328 & 1.64 & 0.44 & 1.03\\
2012-05-23 & J1125$+$2610 & 1123$+$264 & 2.35 & 0.84 & 0.44\\
2012-10-11 & J1340$-$0137 & 1337$-$013 & 1.62 & 0.23 & 0.79\\
2013-06-27 & J0613$+$1306 & 0611$+$131 & 0.74 & 0.34 & 0.91\\
2013-10-14 & J0211$+$1051 & 0208$+$106 & 0.20 & 0.63 & 0.67\\
2013-10-23 & J2009$+$1318 & 2007$+$131 & 0.61 & 0.18 & 1.89\\
2013-12-04 & J1912$-$1504 & 1909$-$151 & 1.80 & 0.31 & 1.36\\
2013-12-04 & J1916$-$1519 & 1914$-$154 &      & 0.18 & 1.07\\
2014-01-08 & J2258$+$0203 & 2256$+$017 & 2.67 & 0.18 & 0.53\\
2014-02-03 & J2327$-$1447 & 2325$-$150 & 2.46 & 0.52 & 2.58\\
2015-08-12 & J2148$+$0657 & 2145$+$067 & 1.00 & 6.60 & 1.38\\
2015-08-12 & J2151$+$0709 & 2149$+$069 & 1.36 & 0.89 & 1.00\\
2015-08-12 & J2151$+$0552 & 2149$+$056 & 0.74 & 0.61 & 0.44\\
2015-08-31 & J0336$+$3218 & 0333$+$321 & 1.26 & 1.63 & 1.76\\
2015-09-04 & J0852$+$2833 & 0849$+$287 & 1.28 & 0.34 & 1.03\\
2015-09-26 & J1256$-$0547 & 1253$-$055 & 0.54 & 15.38 & 1.52\\
2015-11-14 & J0502$+$1338 & 0459$+$135 & 0.35 & 0.64 & 1.22\\
2016-01-28 & J1733$-$1304 & 1730$-$130 & 0.90 & 4.09 & 1.71\\
2016-01-28 & J1738$-$1503 & 1735$-$150 &      & 0.18 & 1.14\\
2016-03-31 & J0105$+$1553 & 0103$+$156 &      & 0.20 & 0.88\\
2016-05-10 & J2326$+$0112 & 2323$+$009 & 1.60 & 0.15 & 1.15\\
2017-03-21 & J0630$-$1323 & 0628$-$133 & 1.02 & 0.35 & 1.72\\
2017-03-21 & J0631$-$1410 & 0629$-$141 & 1.02 & 0.53 & 0.96\\
2017-09-22 & J0509$+$0541 & 0506$+$056 & 0.34 & 0.45 & 0.08\\
2017-11-06 & J2238$+$0724 & 2235$+$071 & 1.01 & 0.21 & 0.45\\
2018-09-08 & J0945$-$0153 & 0943$-$016 & 2.27 & 0.26 & 1.87\\
2019-07-30 & J1503$+$0917 & 1500$+$094 &      & 0.18 & 1.17\\
2019-07-30 & J1504$+$1029 & 1502$+$106 & 1.84 & 1.55 & 0.31\\
2020-11-30 & J0201$-$1132 & 0159$-$117 & 0.67 & 0.65 & 0.56\\
2020-11-30 & J0206$-$1150 & 0203$-$120 & 1.66 & 0.35 & 1.08\\
2021-08-11 & J1803$+$2521 & 1801$+$253 &      & 0.19 & 0.08\\
2022-02-05 & J1743$-$0350 & 1741$-$038 & 1.05 & 4.02 & 0.84\\
2022-03-03 & J1746$+$1127 & 1744$+$114 &      & 0.19 & 1.04\\
2022-03-03 & J1751$+$0939 & 1749$+$096 & 0.32 & 2.68 & 1.77\\
2022-04-25 & J1752$-$1011 & 1749$-$101 &      & 0.19 & 0.54\\
2022-04-25 & J1758$-$1203 & 1755$-$120 &      & 0.15 & 1.87\\
2022-05-13 & J1449$-$0045 & 1446$-$005 & 0.30 & 0.22 & 1.81\\
2022-05-13 & J1451$-$0127 & 1449$-$012 & 1.33 & 0.22 & 1.09\\
\hline
\end{tabular}
\begin{tablenotes}
\item Notes:
Columns are as follows: (1) is the IceCube event date, same as in \autoref{t:events}; (2) and (3) the blazar name in J2000 and B1950 formats; (4) redshift taken from the NASA/IPAC Extragalactic Database; (5) average VLBI flux density at 8 GHz; (6) angular separation on the sky between the blazar and the corresponding IceCube event.
\item
The blazars from the complete VLBI sample \autoref{s:data_vlbi} which fall within the error regions of the IceCube events (\autoref{s:data_icecube}, \autoref{t:events}) assuming $\Delta = 0.45\degree$ (\autoref{f:pvals}, \autoref{s:stat_flux}).
A full machine-readable version of the table is available online.
\end{tablenotes}
\end{table}

\begin{figure}
\includegraphics[width=1.0\columnwidth,trim=0cm 0cm 0cm -0.1cm]{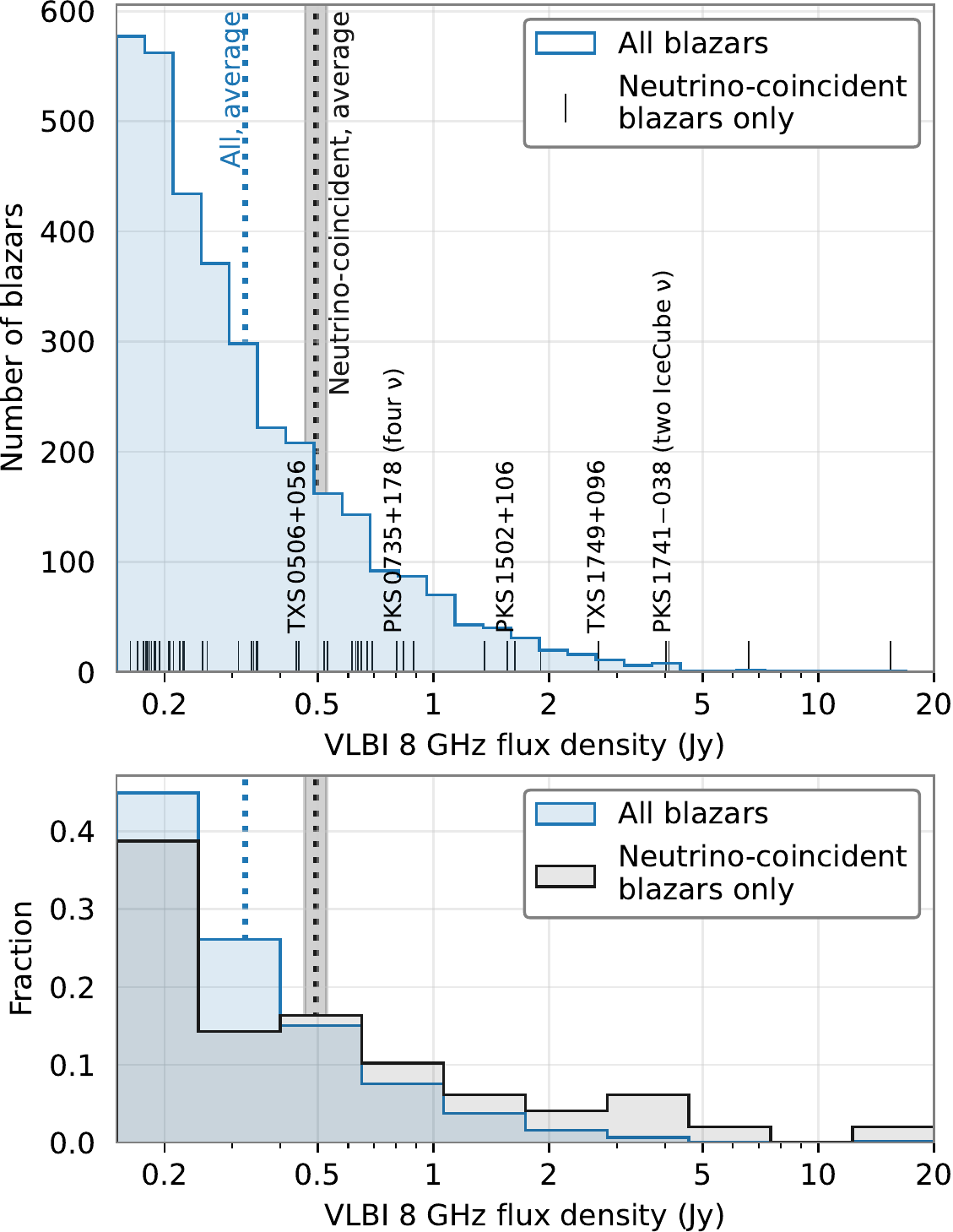}
\caption{
Distribution of the average VLBI flux densities of the blazars. The blue histogram illustrates the complete sample described in \autoref{s:data_vlbi}, with its average value marked. The ticks and the grey histogram indicate flux densities of the blazars falling within the IceCube error regions. Their average is shown together with the shaded 68\% confidence interval. The objects discussed in \autoref{s:individual} are labelled here.
\label{f:avg_fluxes}
}
\end{figure}

Furthermore, we perform this statistical analysis in full on the entire sample of 71 high-energy neutrino events. 
Scanning over $\Delta$ yields a minimum $p$-value of $3\cdot10^{-5}$ when $\Delta=0.45\degree$, the same added error as above; see the 2009-2022 line in \autoref{f:pvals}.
Positional associations between the blazars and high energy neutrinos are listed in \autoref{t:assocs} and visualised in \autoref{f:avg_fluxes}. In case the reported IceCube uncertainties should be interpreted as two-dimensional coverage regions (see Appendix~\ref{a:statdetails}), the optimal $\Delta$ value changes to $0.78\degree$, while the significance level remains consistent.

We conservatively estimate the number of sources driving the correlation following \citetalias{neutradio1}. Specifically, we drop neutrinos coincident with the brightest blazars one by one, and find that the $p$-value exceeds $5\%$ when six brightest matches (6 neutrino -- blazar pairs, 5 blazars) are removed. These matches with brightest blazars include the four selected earlier in \citetalias{neutradio1} (3C~279, NRAO~530, PKS~1741$-$038, PKS~2145+067), and two new events from 2020-2020: the second neutrino from PKS~1741$-$038 (\autoref{s:1741}), and a neutrino coincident with TXS~1749+096 (\autoref{s:J1751}).

The minimum $p$-value is the so-called pre-trial value affected by the multiple comparisons issue. We take this into account using post-trial calculations, as described in Appendix~\ref{a:statdetails}. The post-trial $p$-value is fundamentally unaffected by the multiple tests performed when trying different $\Delta$ values, it takes the shape of the curve in  \autoref{f:pvals} into account. We obtain a final post-trial $p$-value of $3\cdot10^{-4}$. For comparison, the post-trial $p$-value was $2\cdot10^{-3}$ in \citetalias{neutradio1}; this difference illustrates the statistical power gained when adding recent neutrino detections.

The value of the added directional error $\Delta$ is related to systematic uncertainties, but cannot be easily interpreted. The uncertainties are different for different energies, geometries of events, quality cuts and reconstruction procedures. However, the obtained $\Delta=0.45^\circ$ is in agreement with the angular differences between best-fit directions of the same high-energy events in different reconstructions, cf.\ Fig.~2 of \citet{2022PhyU...64.1261T}. Different event reconstructions used in different IceCube studies \cite[e.g.,][]{IceCubeTXSgamma,2021arXiv210708670L,2022Sci...378..538I} can have different error properties that are hard to compare directly. IceCube analysis of directional systematic uncertainties for high-energy alert events, used in our study, is ongoing. Its early results suggest that systematic uncertainties of a fraction of a degree are expected, cf.~Fig.~7 of \citet{2021arXiv210708670L}.

Statistical significances attained in our analyses with different event subsamples are summarised in \autoref{t:pvalues}. 
Supplementing our analysis, we have performed a less direct test with updated event reconstruction from \cite{IceCube:muon2021} that influenced 27 events out of 71. The significant correlation remains after this update when adding the determined $\Delta=0.45\degree$.
\autoref{t:pvalues} also includes an update of our joint analysis of high and lower-energy neutrinos (\citetalias{neutradio2}). For lower energies, we take the IceCube Northern sky map for 2008-2015 \citep{icecubecollaborationAllskyPointsourceIceCube} and correlate it with radio blazars, as was done in \citetalias{neutradio2}. To conservatively ensure the map analysis does not overlap with the same high-energy events, we mask out map pixels close to events from \autoref{t:events} within that time period. Specifically, all pixels within reported event uncertainties and closer than $\Delta=0.45\degree$ are ignored. This lets us combine the high and lower-energy analyses as statistically independent \citep{fisher}, yielding a final p-value of $1.9\cdot10^{-5}$ reported in \autoref{t:pvalues}.

In \citetalias{neutradio1}, we also reported on the detection of temporal correlation between neutrino arrival and radio flares in blazars, most prominent at high frequencies of tens of GHz. Here, we evaluate that correlation using the most up-to-date 71 neutrino detections sample. Adding the 14 recent events does not yield a noticeable increase in the statistical significance compared to the earlier $p$-value of 5\% reported in \citetalias{neutradio1}, which agrees with the considerations of \cite{2022A&A...666A..36L}. The variability analysis necessarily includes fewer blazars than the complete VLBI sample analysis discussed above: only a subset of the objects is regularly monitored. Furthermore, there are more free parameters, including the width of the time window when flares are considered coincident with neutrino arrivals. Therefore, more events and neutrino-blazar associations are required to make reliable, meaningful progress in evaluating temporal correlations.

\begin{figure}
\includegraphics[width=1.0\columnwidth]{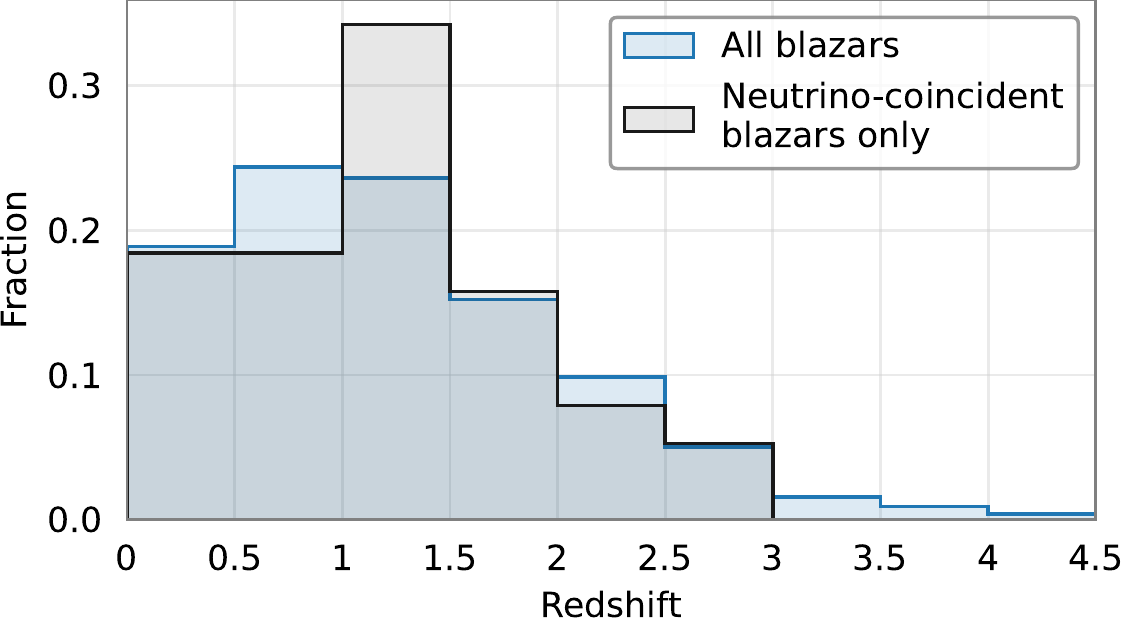}
\caption{ 
The distribution of cosmological redshifts for blazars from the complete VLBI sample (2542 sources), and for blazars coincident with neutrinos (38 sources, see \autoref{t:assocs}). The redshift distribution of neutrino coincidences does not significantly differ from the whole sample, apparent deviations are due to randomness within each bin. Median redshift is between 1.0 to 1.1 for both distributions.
\label{f:redshifts}
}
\end{figure}

\begin{table*}
\caption{
Summary of statistical analyses correlating IceCube neutrinos with radio-bright blazars on the basis of this and previous papers.
}
\label{t:pvalues}
\centering
\begin{tabular}{cp{5cm}ccp{4cm}}
\hline\hline
\# & Dataset & \multicolumn{2}{c}{Statistical significance} & Reference\\
& & p$-$value & Gaussian equivalent & \\
(1) & (2) & (3) & (4) & (5)\\
\hline
\multicolumn{5}{c}{High-energy neutrinos}\\
1 & 56 events $>200$~TeV, 2009-2019 & $2\cdot10^{-3}$ & $3.1\sigma$ & \citetalias{neutradio1} \\
2 & 14 events $\geq200$~TeV, 2020-2022 & $6\cdot10^{-2}$ & $1.9\sigma$ & this paper: new events only \\
3 & 71 events$^\mathrm{a}$ $\geq200$~TeV, 2009-2022 & \textbf{$3\cdot10^{-4}$} & \textbf{$3.6\sigma$} & this paper: all events \\
\hline
\multicolumn{5}{c}{Neutrinos of all energies}\\
4 & 71 high-energy events (\#3) plus lower-energy neutrinos based on the 7-year sky map & \textbf{$1.9\cdot10^{-5}$} & \textbf{$4.3\sigma$} & this paper jointly with the independent \citetalias{neutradio2} analysis \\
\hline
\end{tabular}
\begin{tablenotes}
\item
$^\mathrm{a}$ All high-energy events selected in \autoref{s:data_icecube} and listed in \autoref{t:events}, i.e.\ 2009-2019 collected by \citetalias{neutradio1} and \citetalias{neutradio2}, 2020-2022 from recent IceCube GCN alerts.
\item Notes:
Columns are as follows: 1~--- test number, 2~--- dataset description, 3~--- $p$-value attained in the analysis, 4~--- statistical significance in terms of the two-sided Gaussian equivalent, 5~--- analysis reference.
\item
Analyses differ in the event samples: the number of the detections grows over time, and the datasets with different selection criteria become available. The statistical significances reported here either did not require any trials at all (\#2) or are reported as post-trial probabilities when parameter selection was performed. The most up-to-date statistical results are highlighted in bold.
\end{tablenotes}
\end{table*}

\subsection{Discussion and future prospects} \label{s:tests}

The results presented in \autoref{s:stat_flux} summarise the growing evidence for the neutrino-blazar associations on the basis of recent IceCube events. They include a direct, statistically independent test of our previous findings and predictions \citepalias{neutradio1,neutradio2}. The bright blazars driving the correlation are highlighted in \autoref{f:skymap} and \autoref{f:avg_fluxes}. They are located at cosmological distances $z \sim 1$, corresponding to typical blazar redshifts (\autoref{f:redshifts}; also see \citealt{2021ApJ...923...30L}), and consistent with the space volume distribution.

Since 2020, other works have also noted correlations between blazars and high-energy neutrinos, including studies based on IceCube events \citep{Resconi2020,2021A&A...650A..83H,2022arXiv220313268N,2022ApJ...933L..43B}, results of ANTARES \citep{Illuminati2021, RadioPS} and Baikal-GVD \citep{Baikal2021}.
Most recently, \citet{IceCubeTeVPA2022} have qualitatively confirmed the findings of \citetalias{neutradio1}; they additionally extended the neutrino sample to more than 200 events using different criteria and noted a drop in the correlation significance. These results further indicate the importance of selection criteria and of taking systematic errors into account.

Certain papers did not detect the blazar-neutrino correlation \citep[e.g.][]{2021PhRvD.103l3018Z}. However, their quantitative results are still consistent with our findings. Indeed, \cite{2021PhRvD.103l3018Z} estimate that at most 30\% of high-energy astrophysical neutrino flux comes from radio blazars, while we reported \citepalias{neutradio1} that the fraction is at least 25\%. Still, a couple of possible reasons  could explain why no correlation was detected by \cite{2021PhRvD.103l3018Z}. First, that analysis relies on Gaussian directional errors, while the actual shape is different and has heavier tails \citep{2021arXiv210109836I}; also, systematic uncertainties are not taken into account. The corresponding IceCube 10-year catalogue significantly changed high-energy event directions and errors compared to the earlier data releases. Potential issues are acknowledged in \cite{2021arXiv210109836I}, and the accompanying guidance points out that there are no a priori reasons to prefer the updated sample over the earlier datasets. We provide a more detailed discussion of this and other independent tests in \cite{2022icrc.confE.967P}.

Comparison of $p$-values reported in \citetalias{neutradio1} and \autoref{s:stat_flux} (see also \autoref{t:pvalues}) demonstrates that the significance in terms of $\sigma$s generally follows the $\sqrt{\mathrm{sample\ size}}$ law. 
This also motivates an educated speculation regarding the future prospects of detecting the directional correlation between blazars and high-energy neutrinos. Extrapolating the square root law shows that 150 events would be enough to obtain a $6\sigma$ pre-trial significance, which should lead to a post-trial detection exceeding $5\sigma$. Assuming that IceCube capabilities remain the same, and KM3NeT and Baikal-GVD together provide a similar number of track events of the same quality as IceCube, we expect that 150 high-energy events satisfying our criteria are accumulated within the next five to seven years. Any improvements on this scenario would bring the expected result sooner.

\section{Selected individual blazar-neutrino associations} \label{s:individual}

\begin{figure*}
\newcommand{\height}{0.22\columnwidth}
\newcommand{\width}{0.9\textwidth}
\begin{subfigure}[b]{\width{}}
 \centering
 \includegraphics[height=\height{}]{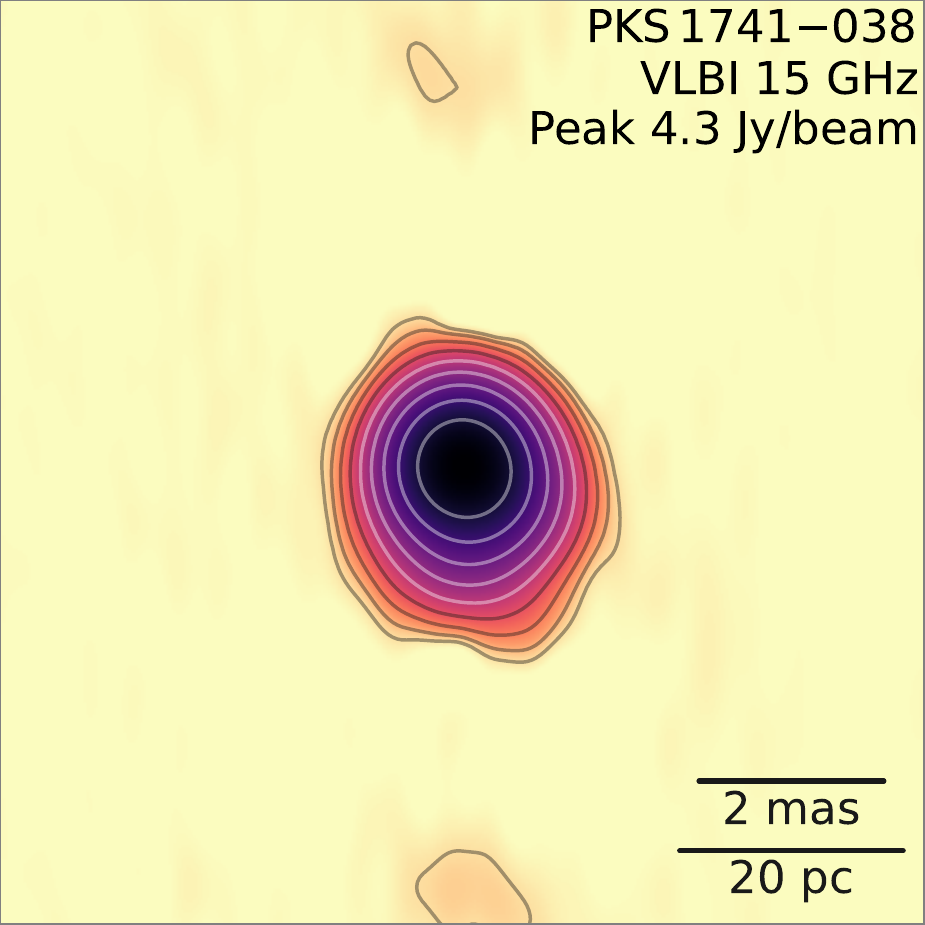}
 \includegraphics[height=\height{}]{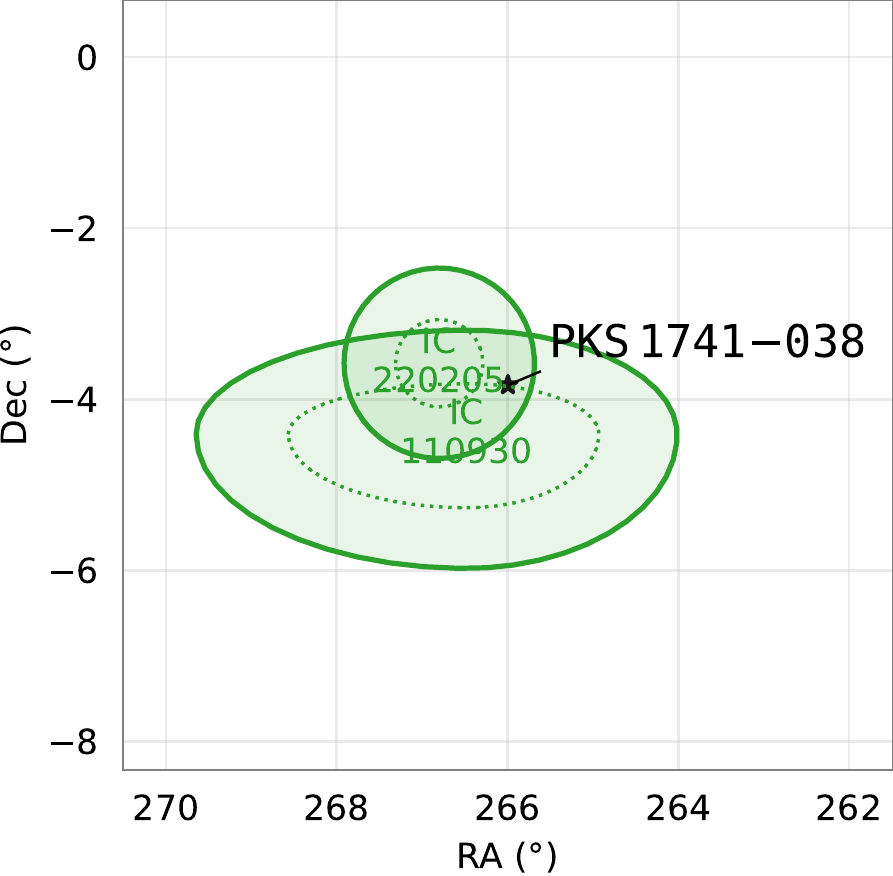}
 \includegraphics[height=\height{}]{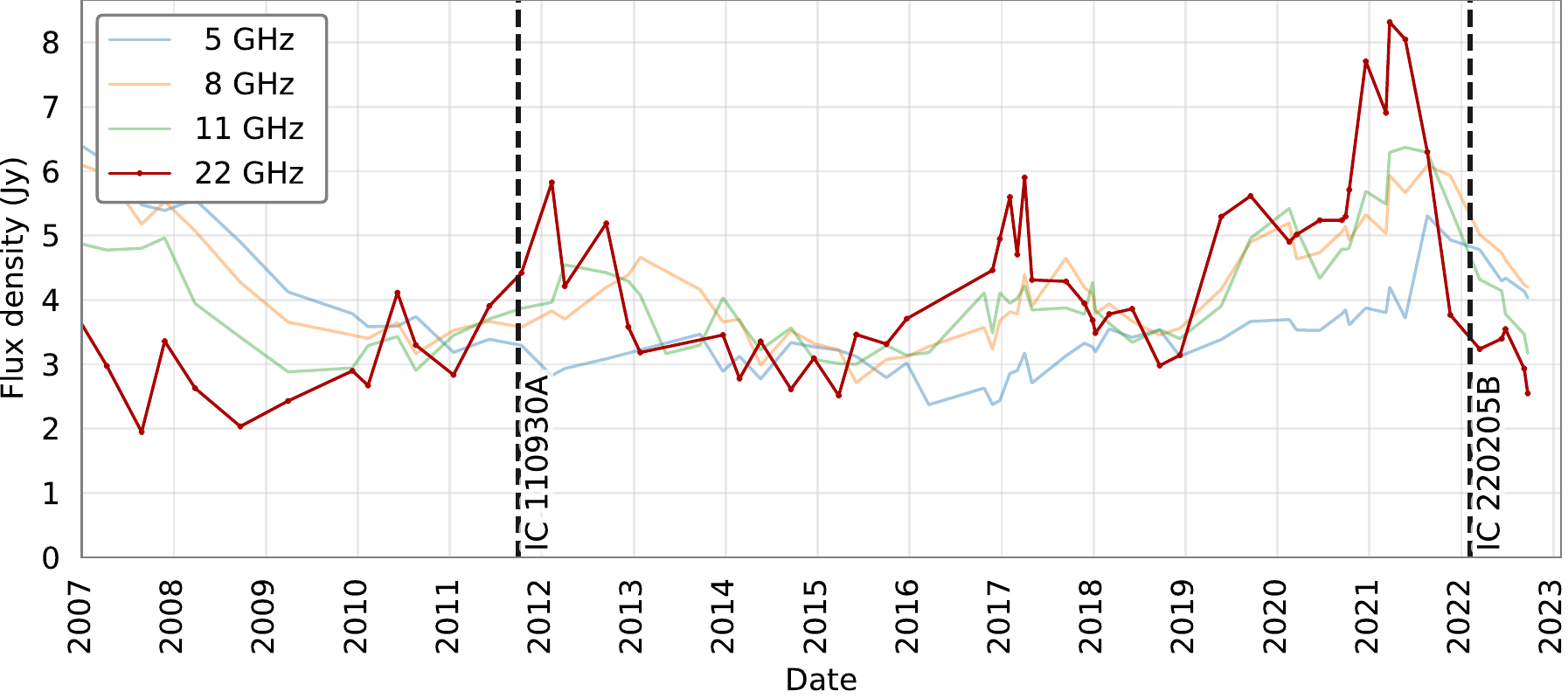}
 \caption{PKS~1741$-$038: a likely source of the doublet among high-energy neutrino events selected in \autoref{s:data_icecube}. This source is one of the brightest blazars in the sky with 4~Jy average VLBI flux density at 8~GHz. The light curves indicate that both the 2011 and the 2022 events were detected close to major radio flares.}
 \label{f:J1743}
\end{subfigure}
\begin{subfigure}[b]{\width{}}
 \centering
\includegraphics[height=\height{}]{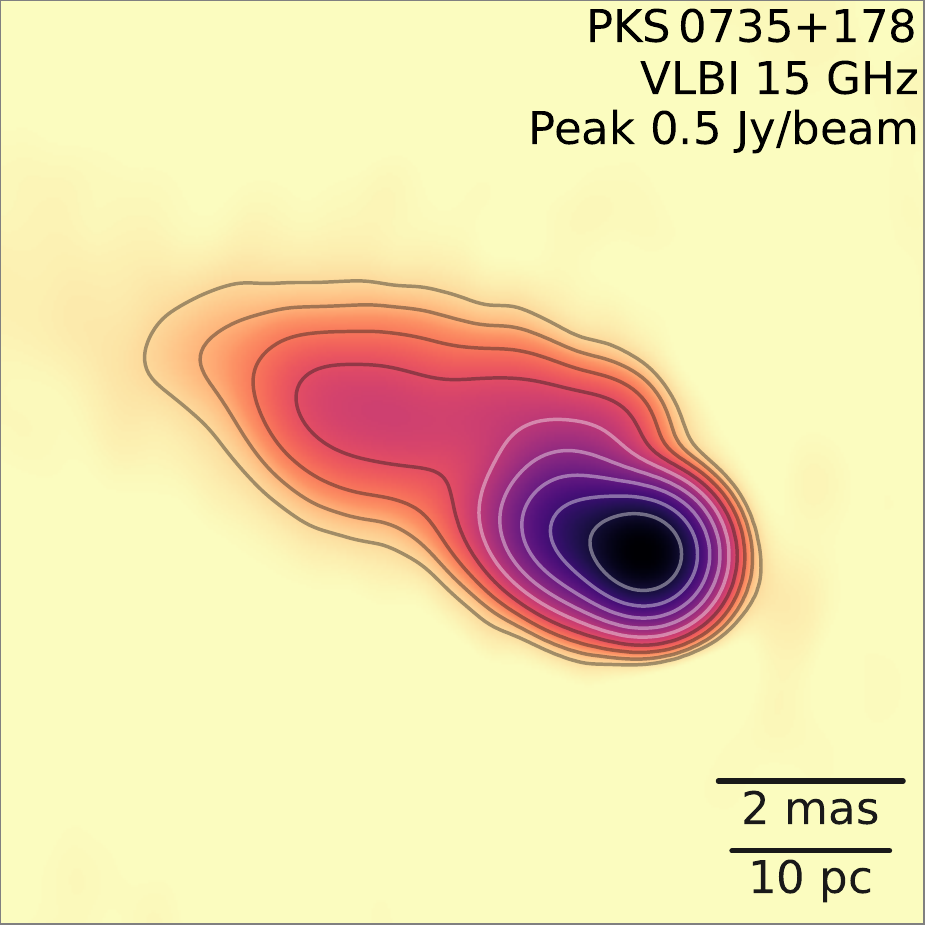}
\includegraphics[height=\height{}]{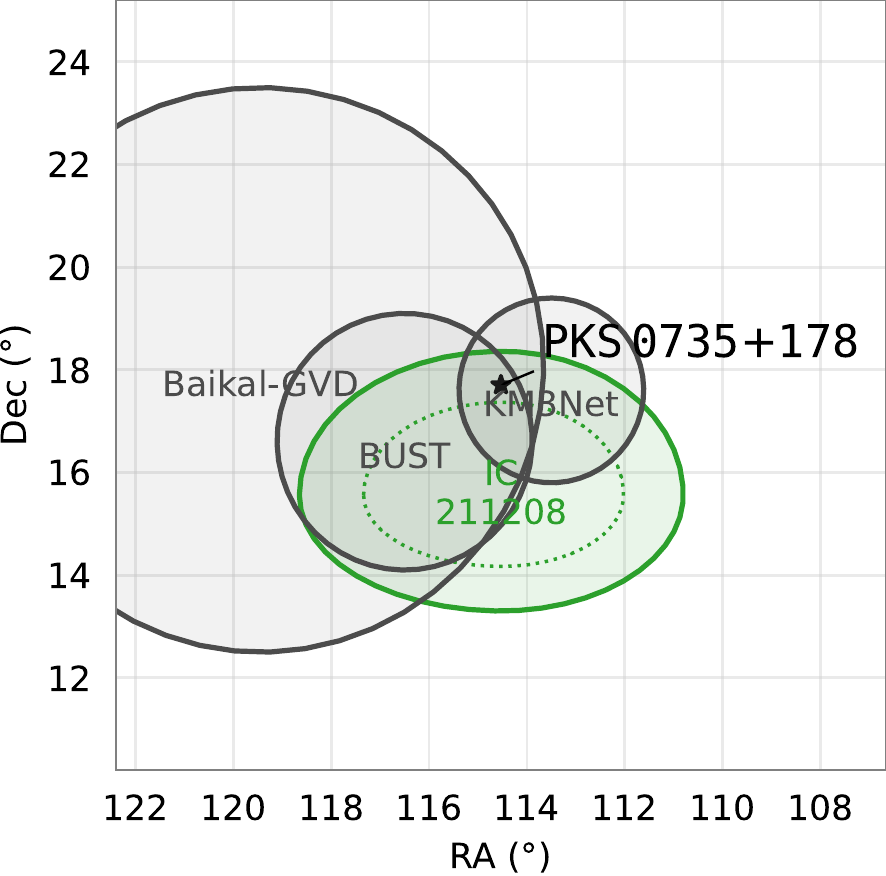}
\includegraphics[height=\height{}]{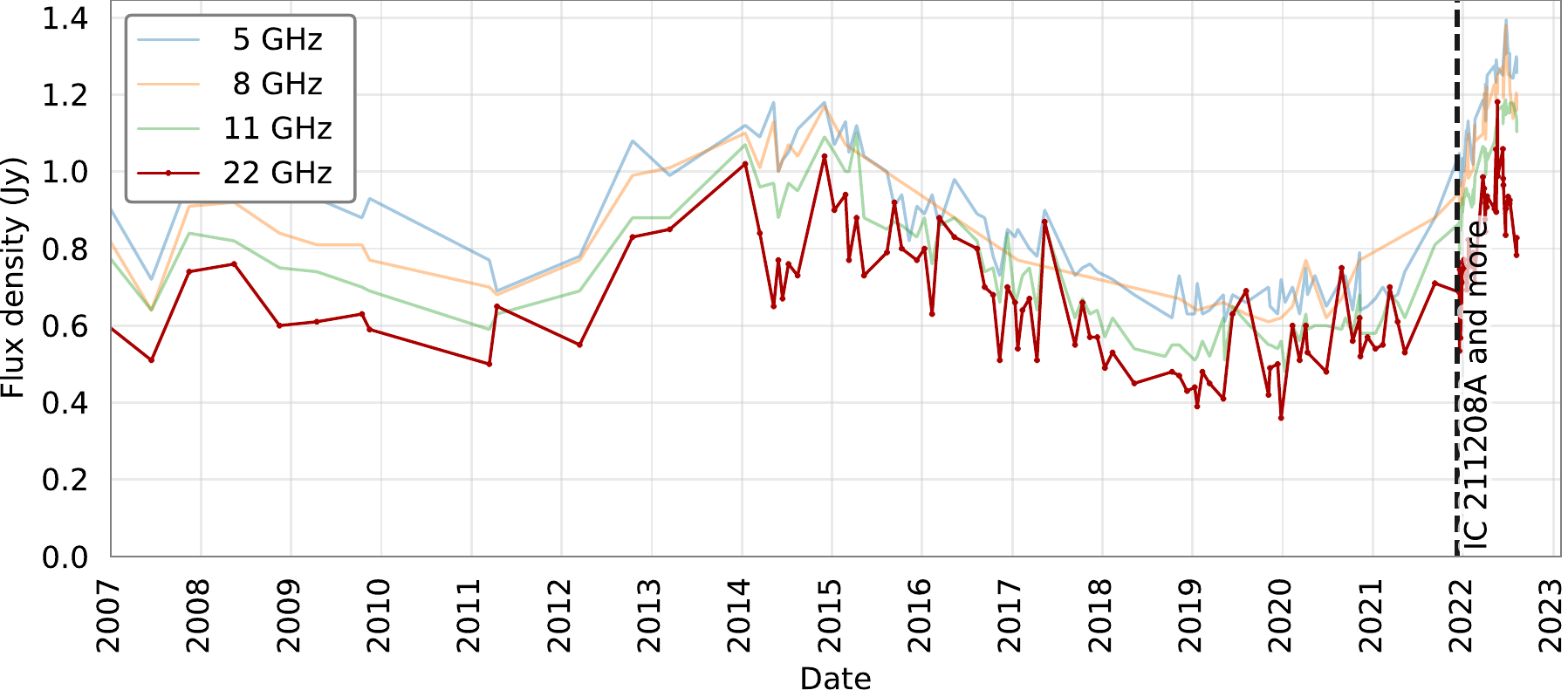}
 \caption{PKS~0735+178: a bright blazar with a major radio flare coinciding with the 2021 IceCube neutrino. The neutrinos detected in December~2021 by Baikal-GVD, KM3NeT and Baksan (BUST) observatories are also shown in the plot with their 50\% containment regions.}
 \label{f:J0738}
\end{subfigure}
\begin{subfigure}[b]{\width{}}
 \centering
\includegraphics[height=\height{}]{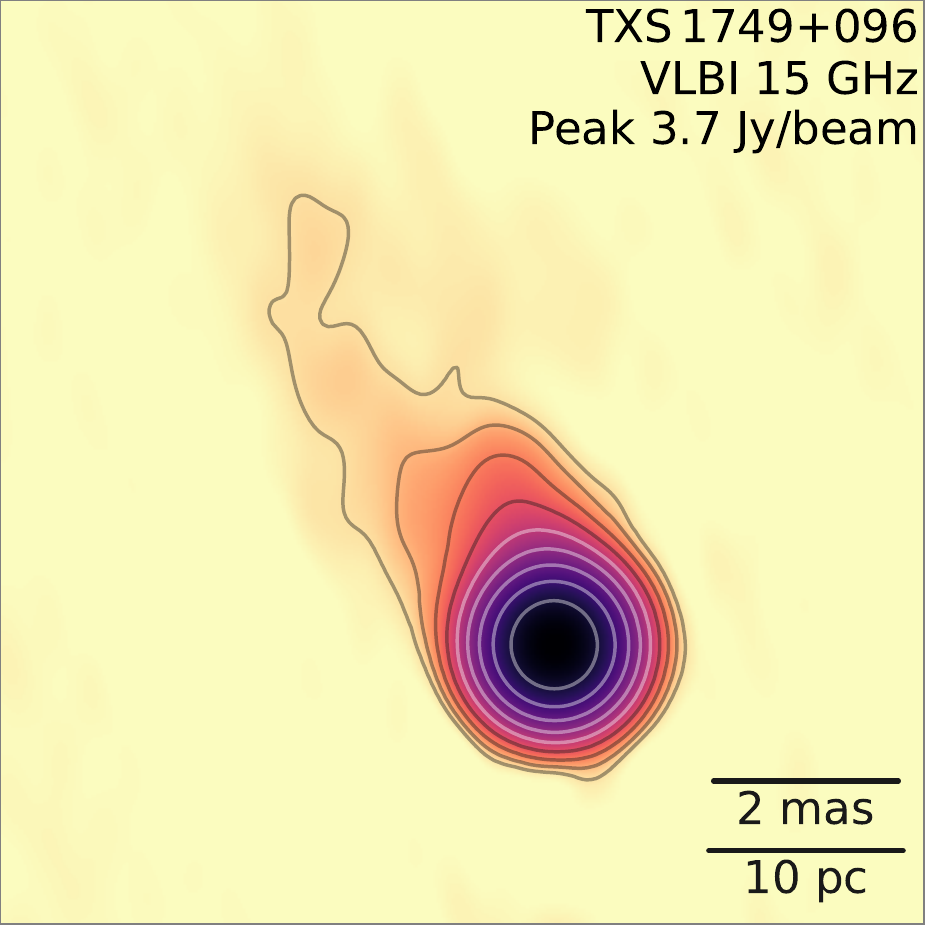}
\includegraphics[height=\height{}]{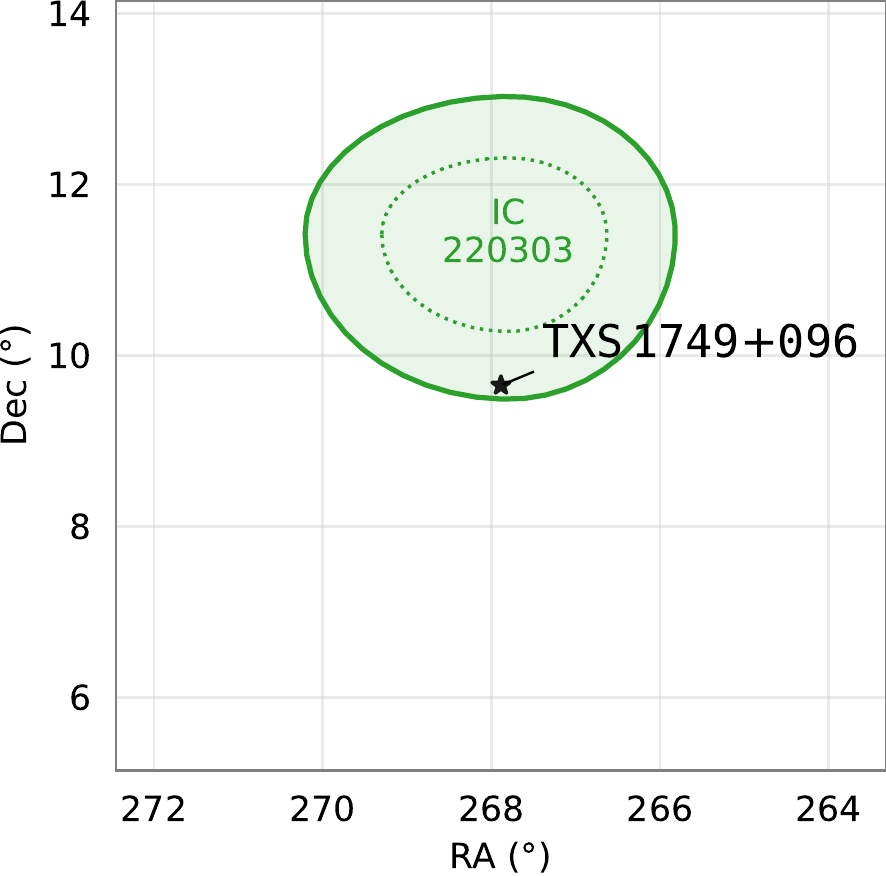}
\includegraphics[height=\height{}]{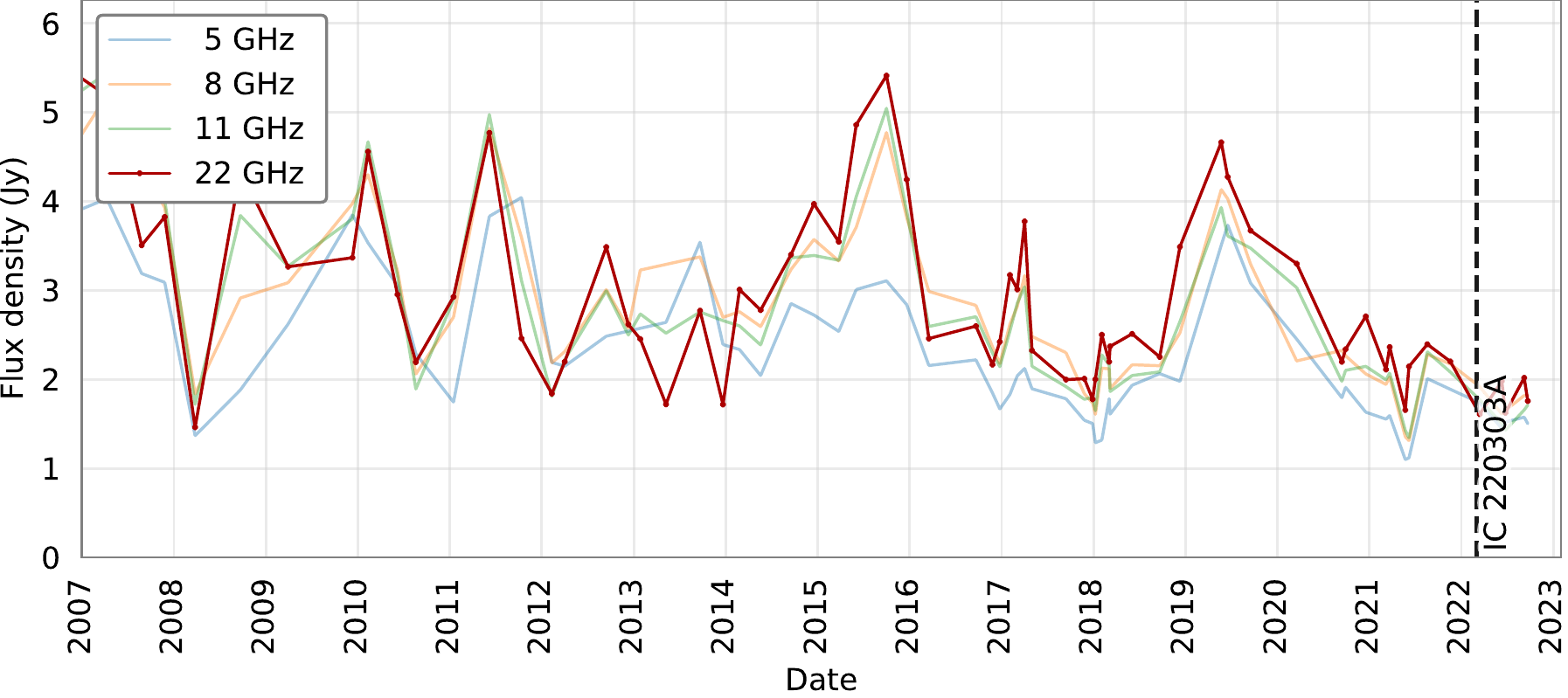}
 \caption{TXS~1749+096: the brightest among the blazars newly matched with the 2020-2022 IceCube events. Its 2.7~Jy average VLBI flux and strong variability indicate high Doppler boosting, discussed in \autoref{s:J1751}.}
 \label{f:J1751}
\end{subfigure}
\begin{subfigure}[b]{\width{}}
 \centering
\includegraphics[height=\height{}]{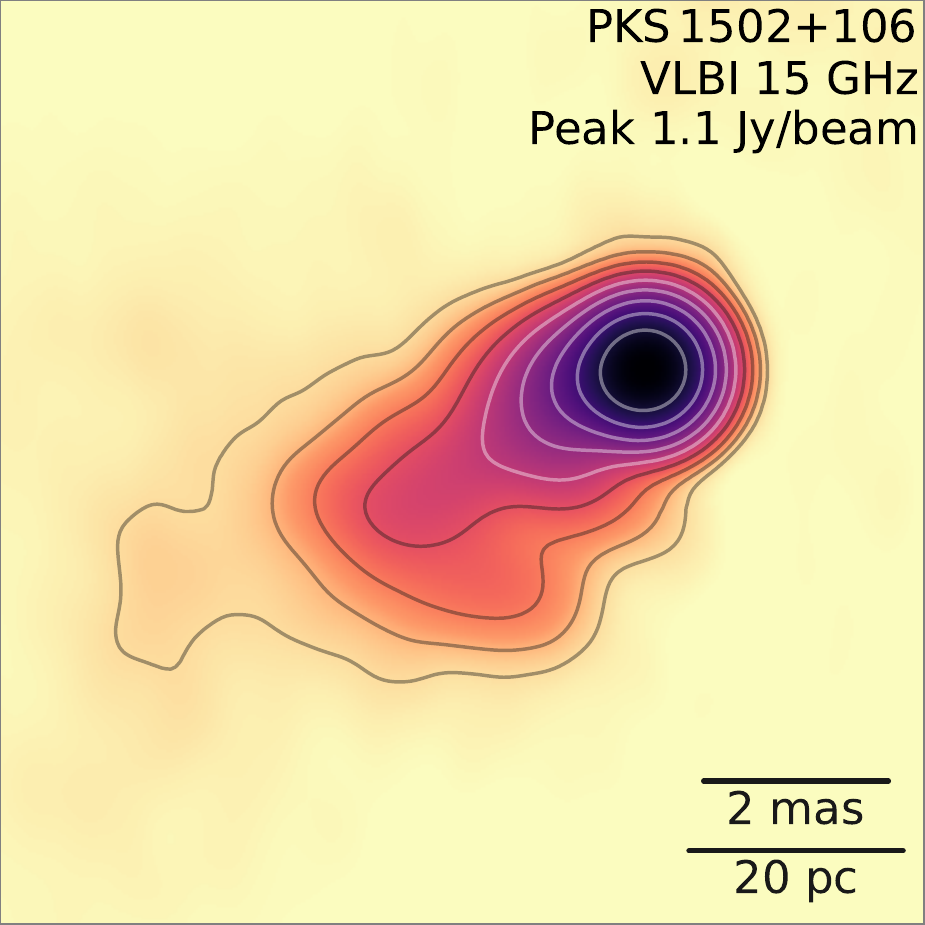}
\includegraphics[height=\height{}]{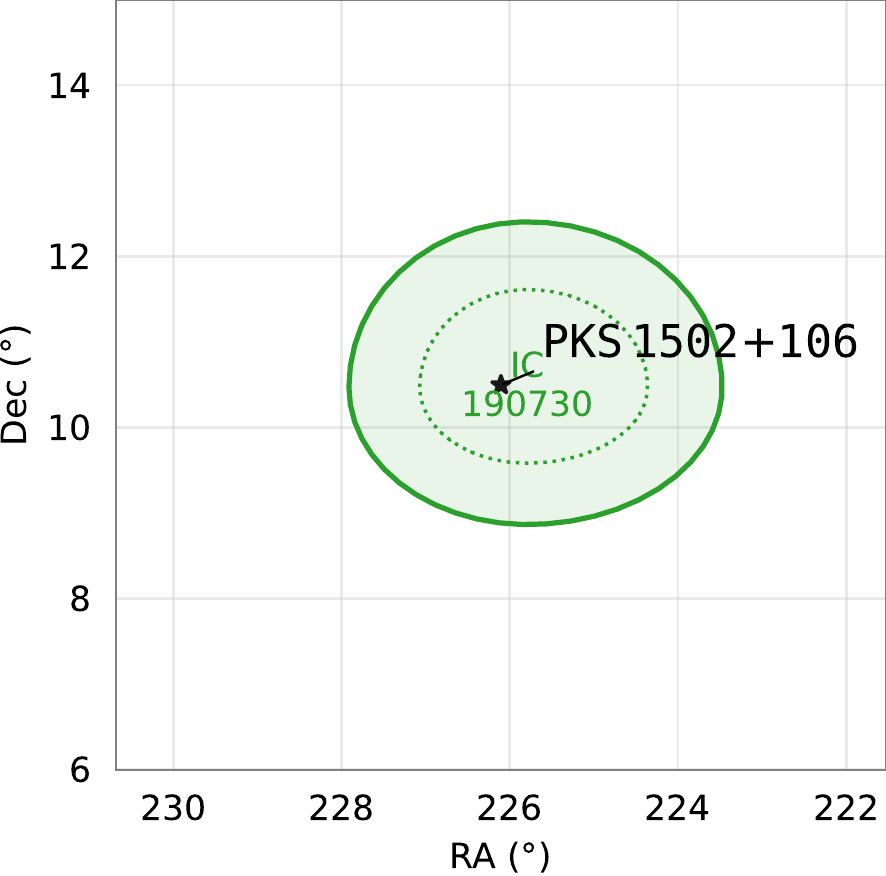}
\includegraphics[height=\height{}]{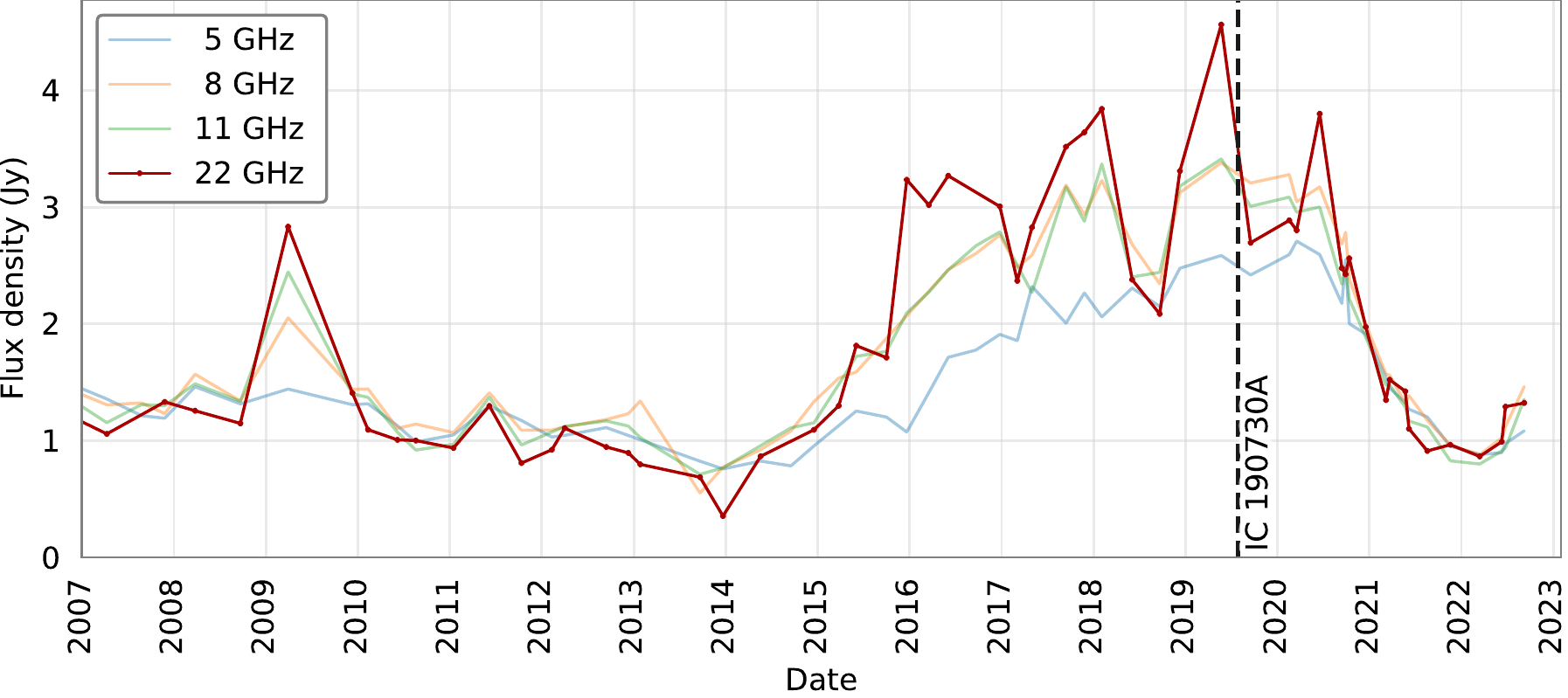}
 \caption{PKS~1502+106: selected for the highest temporal correlation between the neutrino detections and the radio flares in \citetalias{neutradio1}. Later monitoring confirms the correlation and the major flare that receded in 2021-2022.}
 \label{f:J1504}
\end{subfigure}
\begin{subfigure}[b]{\width{}}
 \centering
\includegraphics[height=\height{}]{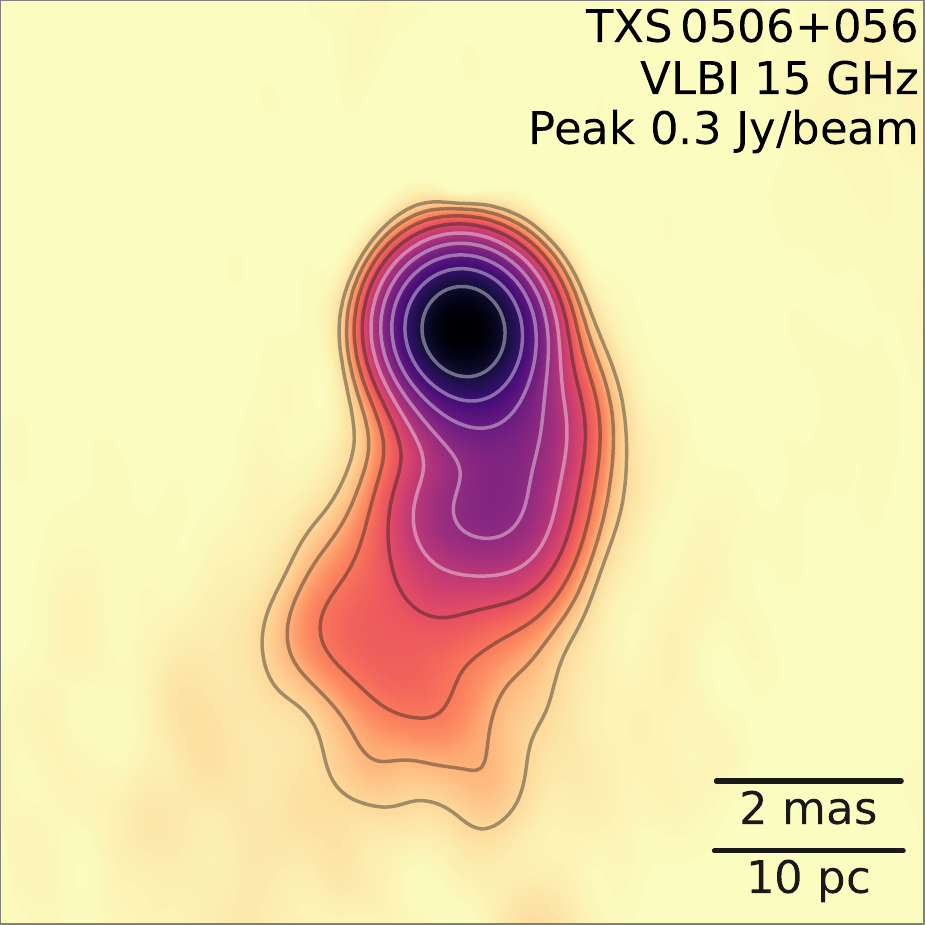}
\includegraphics[height=\height{}]{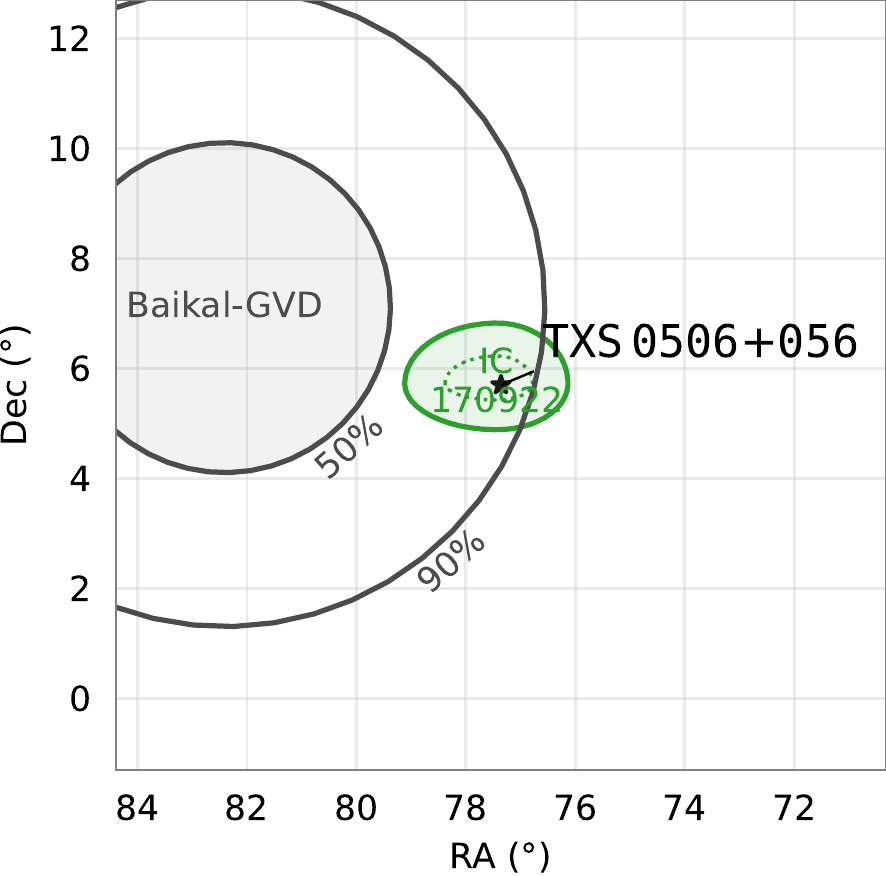}
\includegraphics[height=\height{}]{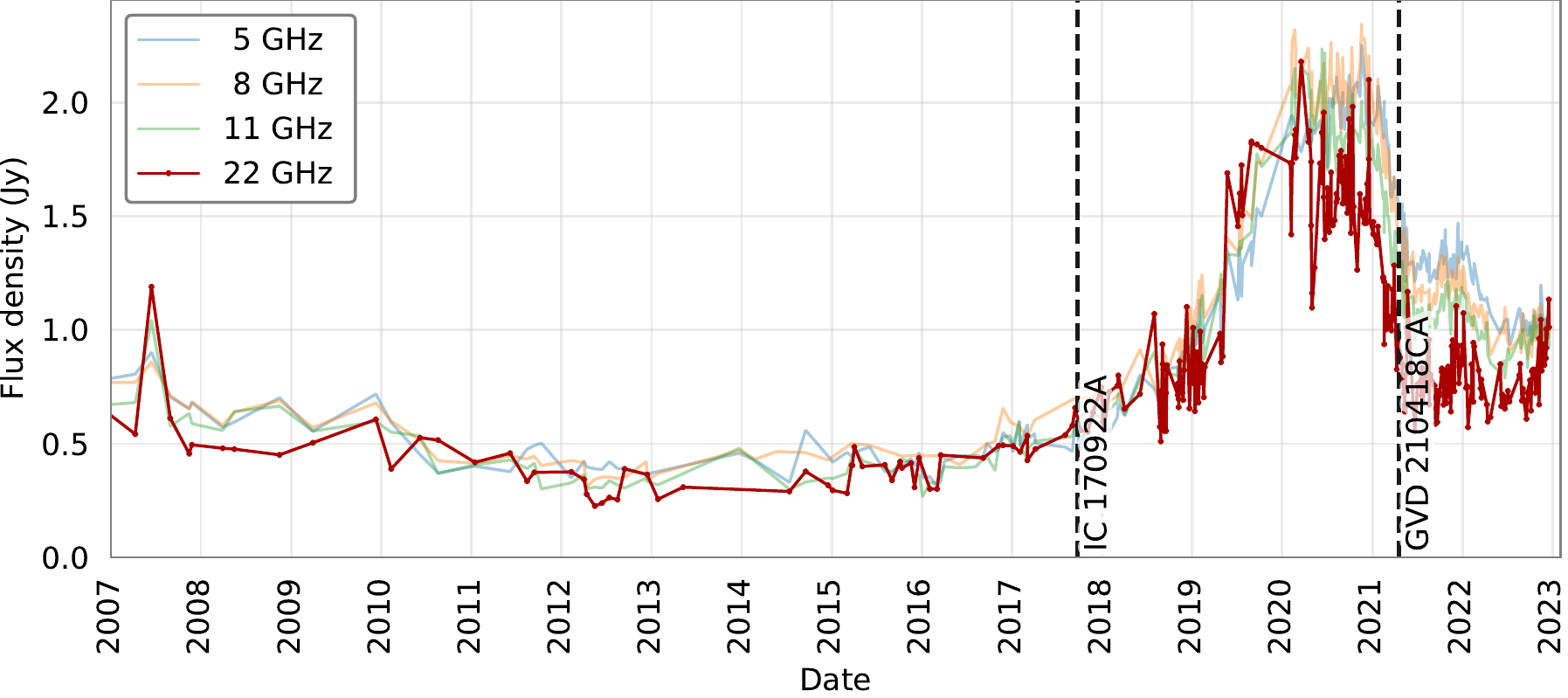}
 \caption{TXS~0506+056: the first high-energy neutrino association. Later monitoring demonstrated a coincident major radio flare.
 The uncertainty ellipse and the arrival time of the recent Baikal-GVD neutrino detection \citep{2022arXiv221001650B} is also shown.
 }
 \label{f:J0509}
\end{subfigure}
\caption{Individual neutrino-blazar associations discussed in \autoref{s:individual}. We show the corresponding sky regions, time-averaged VLBI images from the MOJAVE programme \citep{2017MNRAS.468.4992P} and RATAN-600 radio light curves since 2007 for each association. The green ellipses represent the error regions of the IceCube events enlarged by $\Delta=0.45\degree$ to account for systematic errors (\autoref{s:stat_flux} and Appendix~\ref{a:statdetails}); the dashed green lines indicate the original reported errors at 90\% confidence level.}
\end{figure*}

The statistical analysis conducted in \autoref{s:stat_flux} brings even more evidence for bright blazars spatially correlating with IceCube neutrinos. This section highlights individual associations which we find the most interesting. Neutrinos are considered coincident with blazars when they fall within the IceCube error region expanded by $\Delta=0.45\degree$ according to our analysis in \autoref{s:stat_flux} and Appendix~\ref{a:statdetails}. We focus on the 2020-2022 IceCube events, but also mention earlier relevant detections and reports by other neutrino observatories. We supplement neutrino and VLBI observational results with our RATAN-600 multi-frequency radio light curves \citep{1979S&T....57..324K,1999A&AS..139..545K,2002PASA...19...83K,r:kovalev0506}. Likely associations with earlier detections were reported in \citetalias{neutradio1}. In addition to the individual figures below, these blazars are highlighted and labelled in \autoref{f:skymap}.

\subsection{PKS~1741$-$038: IceCube neutrino doublet}
\label{s:1741}

PKS~1741$-$038 is a bright blazar with an average compact emission at the 4-Jy level from 1 to 100~GHz. It is one of the four blazars we selected in \citetalias{neutradio1} as the most likely neutrino associations: a coincident high-energy IceCube event was detected in 2011. Later, in 2022, IceCube detected another neutrino \citep{2022GCN.31554....1I} coincident with PKS~1741$-$038 \citep{2022ATel15215....1K}, see \autoref{f:J1743}.
We find that the chance coincidence probability for at least one of the IceCube events since 2020 to coincide with one of the four blazars we selected in \citetalias{neutradio1} is 3.5\%, confirming the early ATel report \citep{2022ATel15215....1K}. For a statistical analysis that takes all events and blazars into account, see \autoref{s:stat_flux}.

The light curve (\autoref{f:J1743}) shows that PKS~1741$-$038 is experiencing a major radio flare which peaked at an 8~Jy level in 2021. Another radio flare has occurred in 2011-2012, coincidently with the previous detected neutrino from this direction. Following \citetalias{neutradio1} and \citet{2021A&A...650A..83H}, this temporal correlation forms further support for PKS~1741$-$038 being the source of the two coincident high-energy neutrinos.

\subsection{PKS~0735+178: multiple neutrinos, a major flare}
\label{s:J0738}

On 8~December~2021, IceCube detected a high-energy 171-TeV neutrino track-like event \citep{2021GCN.31191....1I} coincident with the PKS~0735+178 blazar \citep{2021ATel15105....1K}. The reported positional uncertainty has an area of 12~sq.~deg., somewhat larger than the threshold of 10~sq.~deg. we use in \autoref{s:data_icecube}. This event was, therefore, not included in our statistical analysis (\autoref{s:stat_flux}). Still, we believe this particular coincidence is likely a real association.

First, neutrino events from the direction of PKS~0735+178 were also detected in December~2021 by other neutrino observatories, see \autoref{f:J0738} for an illustration. Namely:
\begin{itemize}
    \item Baikal-GVD \citep{2021ATel15112....1D} cascade event with an energy of 43 TeV, just four hours after the IceCube event. The blazar is $4.7\degree$ away from the event, within the 50\% containment region of $5.5\degree$. The probability of a chance coincidence was estimated by \citet{2021ATel15112....1D} as $4.8\cdot 10^{-3}$.
    \item Baksan Underground Scintillation Telescope \citep[BUST;][]{2021ATel15143....1P} event from 4~December~2021, with the blazar $2.2\degree$ away and within the 50\% containment region of $2.5\degree$. It is difficult to determine the neutrino energy given the nature of the instrument: the lower bound is 1~GeV, while the actual energy can be orders of magnitude larger \citep{Petkov:20229B}. Since the event comes from below the horizon, the atmospheric muon background was suppressed and the probability of a chance coincidence with the blazar flare is $2.5\cdot 10^{-3}$ \citep{2021ATel15143....1P,Petkov:20229B}.
    \item There is some indication (p-value of 0.14) that a KM3NeT neutrino detection on 15~December~2021 \citep{2022ATel15290....1F} is related to this blazar.
\end{itemize}
Preliminary statistical significance estimates for each of these coincidences are reported in the corresponding ATels listed above. Here, we do not attempt to combine these estimates and analyse them together: they are not the results of a uniform statistical analysis.

Second, PKS~0735+178 has experienced a major electromagnetic flare coincident with the neutrino detection. The enhanced emission was seen in the whole spectrum, from radio (\autoref{f:J0738}) through optical \citep{2021ATel15021....1S} and X-ray \citep{2021ATel15102....1S,2021ATel15108....1H} to gamma rays \citep{2021ATel15099....1G}. See, e.g. \cite{2023MNRAS.519.1396S} for modelling of EM and neutrino emission of PKS~0735+178. Following the findings of \citetalias{neutradio1} and \cite{2021A&A...650A..83H}, strong temporal correlation is an extra argument for those neutrinos being associated with the blazar. 

\subsection{TXS~1749+096: the brightest new association in 2020-2022}
\label{s:J1751}

Following the same approach as in \citetalias{neutradio1}, we select blazars with the brightest radio emission among those coincident with neutrinos as the most likely associations. There are two blazars above 1 Jy VLBI flux density close to 14 IceCube alerts since 2020 (\autoref{s:data_icecube}): PKS~1741$-$038 at 4~Jy discussed above and TXS~1749+096 with an average flux density of 2.7 Jy. This blazar coincides with the IC220303A event from March 2022, see \autoref{f:J1751}.

The radio light curve of TXS~1749+096 indicates very high levels of flaring activity all the time: the blazar experiences major radio flares every year or two. Together with bright, compact parsec-scale emission, this is an indication of a high Doppler boosting. Among 447 AGNs in the MOJAVE (Monitoring Of Jets in Active galactic nuclei with VLBA Experiments) programme \citep{2018ApJS..234...12L,2021ApJ...923...30L}, TXS~1749+096 demonstrates the third-highest Doppler factor that exceeds 100 \citep{2021ApJ...923...67H}. This fits our interpretation that the radio blazar-neutrino correlation is predominantly driven by beaming effects \citepalias{neutradio1,neutradio2}. Relativistic beaming may affect neutrinos and electromagnetic emission differently, and comparing these effects may help further constraining emission models \citep[see models by ][]{Mannheim1,NeronovWhich}.

\subsection{PKS~1502+106}

PKS~1502+106 experienced a major flare coincident with a neutrino detection in 2019, visible from radio \citep{r:kielmann1502ATel} to gamma rays \citep{2021ApJ...912...54R}. Earlier, \citetalias{neutradio1} selected PKS~1502+106 as a likely neutrino source: it is the blazar with the strongest temporal correlation between radio emission and IceCube neutrino arrival. Following the reports from the ATel \citep{r:kielmann1502ATel} and our selection of this blazar as a probable neutrino association, PKS~1502+106 became an object of significant astrophysical interest. Several works present models describing its electromagnetic and neutrino emission, see, e.g. \cite{2021ApJ...912...54R,2021JCAP...10..082O,2022evlb.confE...9S,2022aems.conf..284K,2022muto.confE..27K}. Later observations confirm a strong temporal correlation due to the major radio flare in 2016-2020, which has currently receded, see \autoref{f:J1504}.

\subsection{TXS~0506+056}

The TXS~0506+056 blazar was the first to be associated with a high-energy neutrino event \citep{IceCubeTXSgamma}. Numerous later studies focused on different aspects of TXS~0506+056 from observational and theoretical perspectives \citep[e.g.][]{MuraseTXS,r:ros0506,r:kovalev0506}. Based on the observational properties, it was argued that the source is a typical radio-bright blazar \citep{r:kovalev0506}. Neutrino-blazar associations presented in \citetalias{neutradio1}, \citet{2021A&A...650A..83H}, and in this work share defining characteristics with TXS~0506+056. Those blazars tend to demonstrate bright radio emission from parsec scales and be highly variable. Flares often coincide with neutrino detections, which is also the case with TXS~0506+056, see light curves in \autoref{f:J0509}. This context brings further evidence for TXS~0506+056 being an ordinary neutrino-emitting blazar. This object is not distinguished in our analysis of bright blazars due to its average long-term VLBI flux density being relatively low, at the $0.4$-$0.5$~Jy level. The flux increased by a factor of 4-5 during the recent flare (see the RATAN-600 light curve in \citealt{2022arXiv221001650B}), but this is not reflected yet in the VLBI catalogue.

In addition to the 290 TeV event in 2017, IceCube reported lower-energy neutrinos detected in 2014-2015 \citep{IceCubeTXSold}. Furthermore, the Baikal-GVD observatory detected a 224~TeV cascade from the direction of TXS~0506+056 in 2021; see \citet{2022arXiv221001650B} for the evaluation of this coincidence and discussion of its implications.

\section{Summary}

We test our earlier results and predictions on the neutrino-blazar connection \citepalias{neutradio1}, utilising recent IceCube events detected in 2020-2022. The updated statistical analysis of 2009-2022 high-energy neutrinos is in agreement with expectations, supporting that the original result was not a statistical fluctuation.
The growing sample of neutrinos brings more evidence for the correlation, which is now detected with a post-trial $p$-value of $p=3\cdot10^{-4}$ ($3.6\sigma$). Blazars with compact parsec-scale radio emission drive this correlation. 
We confirm the importance of taking systematic errors into account and refine the optimal value which should be added to IceCube uncertainties from $\Delta=0.5\degree$ in \citetalias{neutradio1} to $\Delta=0.45\degree$ in this paper ($\Delta=0.78\degree$ if reported IceCube uncertainties represent two-dimensional coverage).
Recently, more tests of our \citetalias{neutradio1} findings have appeared. Their results partly confirm our original findings and further indicate the importance of constructing high-quality neutrino and blazar samples and carefully taking systematic errors into account.

Recent years have brought both confirmations for our earlier blazar-neutrino associations, and new promising coincidences. Among the four brightest blazars we selected in \citetalias{neutradio1} as the most probable neutrino sources, PKS~1741$-$038 had a second neutrino detected from its direction in 2022. The highest neutrino-radio flare correlation in 2020 was found for PKS~1502+106, and electromagnetic observations provide more indications for it being a true association in the meantime. Regarding more recent associations, in 2021, an IceCube neutrino was detected coincident with PKS~0735+178, with other neutrino observatories also indicating a signal from that direction. This blazar was experiencing a major electromagnetic flare at that time. These and other notable cases discussed in \autoref{s:individual} indicate that evaluating compact radio emission is an effective way to search for sources of high-energy neutrinos.

We have demonstrated through statistical evidence and individual examples that blazars are prominent high-energy neutrino sources. Based on the strong correlation with VLBI-bright blazars, we can confirm earlier interpretations and expand on them. First and foremost, neutrinos are produced in central parsec-scale regions of bright blazars. Relativistic beaming is important for them to be detected: neutrinos are observed preferentially from Doppler-boosted jets. Assuming the $p\gamma$ production mechanism, protons of multi-PeV energies are required. They should be accelerated closer to the black hole, and gain bulk motion along the jet direction \citep{Kivokurtseva}. Target photons for the interactions should have energies from keV to hundreds of keV \citepalias{neutradio2}. Photons can either be produced in the jet itself through self-Compton scattering that inevitably accompanies synchrotron radio emission \citep{2009herb.book.....D}, or come from external photon fields closer to the central engine \citep{corr-1807.04299Tavecchio}. Despite growing observational evidence, details on neutrino production mechanisms and localization remain unknown. Future studies of astrophysical neutrinos, together with electromagnetic monitoring of promising blazars, are required to determine the mechanism and the site of neutrino production unambiguously.

Neutrino observatories are growing: Baikal-GVD has recently become fully operational \citep{Baikal-Neutrino2022a}, KM3NeT will join soon \citep{2019APh...111..100A}. This would bring more neutrino detections with a better positional accuracy because of the differences between water and ice. We expect that the neutrino-blazar correlation significance would exceed the $5\sigma$ level within the next five-seven years if following the same statistical approach. Numerous observational campaigns at radio telescopes specifically aim at finding and understanding neutrino-associated blazars better. These include the MOJAVE\footnote{\url{https://www.cv.nrao.edu/MOJAVE/}}
and Boston University\footnote{\url{https://www.bu.edu/blazars/BEAM-ME.html}} VLBA monitoring programmes; improving the completeness and depth of the VLBI-selected blazar sample further (VLBA project BP252); our ongoing neutrino-triggered VLBA (BK232, BK247) and RATAN-600 programmes.

Detections of neutrinos emitted by bright blazars do not preclude the presence of other significant components in the astrophysical neutrino flux. Multiple components can help solve the current tensions with gamma-ray measurements \citep[e.g.][]{IceCube-HESE-2020}. In particular, there appears to be increasing evidence for a Galactic neutrino contribution \citep{IceCube:cascades-Gal2sigma,neutgalaxy,ANTARES-GalRidge}. Potentially, there may be no single dominant origin of high-energy astrophysical neutrinos, making the neutrino sky very diverse.

\section*{Acknowledgements}

We thank Anna Franckowiak and Cristina Lagunas Gualda for the discussion of the statistical analysis and directional errors, Eduardo Ros, Egor Podlesnyi and the anonymous referee for helpful comments and discussions on various parts of this work.
We are grateful to Elena Bazanova for language editing.
This work is supported in the framework of the State project ``Science'' by the Ministry of Science and Higher Education of the Russian Federation under the contract 075-15-2020-778.
This research made use of data from the MOJAVE database maintained by the MOJAVE team \citep{2018ApJS..234...12L}.
This research made use of NASA's Astrophysics Data System.

\section*{Data Availability}

The analysis is based on the VLBI observations compiled in the Astrogeo\footnote{\url{ http://astrogeo.org/vlbi_images/}} database and the Radio Fundamental Catalogue\footnote{\url{http://astrogeo.org/sol/rfc/rfc_2022b/}}. The IceCube neutrino detections are collected from the published sources as described in \autoref{s:data_icecube}.
The RATAN-600 data used in \autoref{s:individual} are available from the corresponding author upon reasonable request.

\bibliographystyle{mnras}
\bibliography{neutradio2022}

\appendix
\section{Calculating the statistical significance}
\label{a:statdetails}

The performed statistical procedures are motivated and briefly described in \autoref{s:stat_flux}. We appreciate and even hope that readers may want to fully reproduce our analysis and compare numerical $p$-values. To facilitate this, we provide not only the data being used (see \autoref{s:data}) but also a step-by-step recipe here. It is fully consistent with the analysis performed in \citetalias{neutradio1} and in this paper. Here, we detail the algorithm for readers' convenience and for the ease of reproducibility.

Our aim is to determine whether neutrino-emitting blazars tend to be stronger than average in terms of their radio emission from compact parsec-scale central regions. We use the average historic VLBI flux density of AGNs taken from the RFC catalogue (\autoref{s:data_vlbi}) to evaluate the bright blazar-neutrino correlation significance. The catalogue is continuously updated, with specific versions linked in each paper; the measured flux densities may change. However, the differences are minor and not noticeable in our analysis.

First, we define what it means for blazars to be considered coincident with a neutrino. We interpret reported uncertainties of IceCube events as 90\% coordinate-wise intervals in Right Ascension and Declination, and transform them to obtain two-dimensional 90\% coverage regions. Specifically, we multiply the coordinate-wise errors by the ratio of 90\% quantiles of two- and one-dimensional Gaussian distributions: $\displaystyle\frac{\sqrt{-\log{(1-0.9)}}}{\mathrm{erf}^{-1}(0.9)} \approx 1.30$. We check that all our statistical results remain consistent without this multiplication as well, to account for the potential scenario that reported IceCube uncertainties already correspond to two-dimensional coverage. Either way, error regions are bounded by four quarters of ellipses, as IceCube reports two-sided uncertainties for each coordinate. Next, we need to account for systematic errors in IceCube event directions. These errors are always present, but their values are not available for all events.
For instance, IceCube publishes 90\% point-spread-function containment coordinate uncertainties for its GCN alerts. From \cite{2022arXiv221004930A}, one concludes that these uncertainties include systematic errors estimated by scaling of those obtained for one particular event geometry. However, \citet{2021arXiv210708670L} demonstrated that the systematic uncertainty depends strongly on the geometry and, as a result, errors might be misestimated.
Thus, we introduce the systematic error proxy as a free parameter $\Delta$~--- the same for all events and directions on the sky. Blazars are considered coincident with a neutrino event when they lie inside the 90\% neutrino error ellipse, or outside but closer than $\Delta$ to the nearest point on that ellipse.

Second, we define the test statistic $v$ for the analysis. The main goal of the statistical analysis in this paper is to check the results of \citetalias{neutradio1} with a larger event sample, so we follow the choices made in that paper. Specifically, we set $v$ to the geometric average of 8~GHz VLBI flux densities of all blazars coincident with a neutrino from the sample listed in \autoref{s:data_icecube}. The geometric mean (or, equivalently, the arithmetic mean of logarithms) is used because the range of flux densities can cover several orders of magnitude, and we want to highlight relative differences instead of a few brightest objects dominating the analysis.

To test if the test statistic is significantly higher than could arise by chance, we employ Monte-Carlo simulations in the following way:
\begin{enumerate}
    \item Compute the $v$ statistic using real positions of blazars and IceCube events. Denote its value as $v_\mathrm{real}$.
    \item Repeat $n=10^6$ times the following:\\
        \indent\indent$\bullet$ shift IceCube events to random Right Ascension coordinates, keeping their Declinations and full error regions unchanged;\\
        \indent\indent$\bullet$ compute the $v$ statistic for these randomly shifted events in place of real ones. Denote this value $v_i, \  1 \leq i \leq n$.
    \item The obtained empirical distribution of $v_i$ represents the test statistic distribution under the null hypothesis of no neutrino-blazar spatial correlation.
    \item Count the number $m$ of random realizations with the values $v_i$ not lower than the one for rear realizations $v_\mathrm{real}$: $$m = \#\{i: v_i \geq v_\mathrm{real}\}.$$
    Calculate the raw $p$-value, the probability of a chance coincidence, as $$p^\mathrm{raw} = \displaystyle\frac{m+1}{n+1}$$ following \citet{Davison2013BMA2556084}.
\end{enumerate}

These raw $p$-values depend on the value of the $\Delta$ error term, which is a free parameter. In our analysis, we try multiple $\Delta$ values: from $0\degree$ to $1\degree$ with a step size of $0.01\degree$. This requires correcting for the multiple comparisons issue which we perform with another layer of Monte-Carlo simulations:
\begin{enumerate}
    \item Repeat the following for the real data, obtaining $p^\mathrm{pre}$, and then for IceCube events with randomly scrambled Right Ascensions $N=10^6$ times, obtaining $p_k^\mathrm{pre}, \  1 \leq k \leq N$:\\
        \indent\indent$\bullet$ for each assumed value of the additional error $\Delta$, compute $p^\mathrm{raw}$, as described above;\\
        \indent\indent$\bullet$ take the minimum of these raw $p$-values, which corresponds to the value of $\Delta$ yielding $v$ most significantly different from simulations. This minimum is called the pre-trial $p$-value $p^\mathrm{pre}$.
    \item Count realizations with a lower pre-trial $p$ than the real one: $$M = \#\{k: p_k^\mathrm{pre} \leq p^\mathrm{pre}\}.$$
    Compute the final post-trial $p$-value as the fraction $$p = \displaystyle\frac{M+1}{N+1}.$$
\end{enumerate}
The post-trial $p$-value computed in this way is unaffected by the multiple comparisons issue. This is the final significance estimate reported in the paper.

\bsp	
\label{lastpage}
\end{document}